\documentclass[aps,prb,floatfix,twocolumn,showpacs]{revtex4}

\usepackage{graphicx}

\begin{document}

\title{Optimization of inhomogeneous electron correlation factors 
       in periodic solids}
\author{David Prendergast} 
\author{David Bevan}
\altaffiliation{Present address: Institute of Metallurgy and
          Materials Science, The University of Birmingham, Edgbaston, 
          Birmingham, B15 2TT, United Kingdom.} 
\affiliation{Department of Physics, National University of Ireland, 
             Cork, Ireland.}
\author{Stephen Fahy}
\affiliation{Department of Physics 
             and National Microelectronics Research Centre, 
             National University of Ireland,
             Cork, Ireland.}

\date{\today}

% ---------------------------------------------------------------------- 
%  Abstract 
% ---------------------------------------------------------------------- 

\begin{abstract}
A method is presented for the optimization of one-body and inhomogeneous 
two-body terms in correlated electronic wave functions of Jastrow-Slater 
type.
The most general form of inhomogeneous correlation term which is 
compatible with crystal symmetry is used and the energy is minimized
with respect to all parameters using a rapidly convergent iterative 
approach, based on Monte Carlo sampling of the energy and fitting of energy 
fluctuations.
The energy minimization is performed exactly within statistical sampling 
error for the energy derivatives and the resulting one- and two-body terms 
of the wave function are found to be well-determined.
The largest calculations performed require the optimization of over 
3000 parameters.
The inhomogeneous two-electron correlation terms are calculated for
diamond and rhombohedral graphite. 
The optimal terms in diamond are found to be
approximately homogeneous and isotropic over all ranges of electron separation,
but exhibit some inhomogeneity at short- and intermediate-range, 
whereas those in graphite are found to be homogeneous at short-range,
but inhomogeneous and anisotropic at intermediate- and long-range 
electron separation.
\end{abstract}


\maketitle

% ------------------------------------------------------------------------
\section{Introduction}
\label{Sec.intro}
% ------------------------------------------------------------------------

An accurate description of electron correlation is one of the central
issues in modern electronic structure theory. In principle,
this involves solving the many-electron Schr\"odinger equation,
which for a system of $N$ interacting electrons is an
inseparable  $3N$-dimensional problem.
Methods based on mean-field approximations of the electron interaction,
such as Hartree-Fock (HF) or 
density-functional theory,\ \cite{HohenbergKohn} (DFT) reduce
this to $N$ independent problems, which we may solve to approximate
many physical properties with reasonable accuracy.
However, a more complete description of electron correlation 
is often required to study delicate phenomena such as long-range
electron correlation and van der Waals interactions.

Quantum Monte Carlo (QMC) methods provide an important approach 
for solving the full $3N$ dimensional problem.\ \cite{RMPQMCsolids}
One such method, variational Monte Carlo (VMC), allows us to estimate
expectation values for a given trial wave function. 
Ideally, this wave function is an eigenstate of the many-body Hamiltonian. 
In practice, a parameterized form is used, which approximates the
exact eigenstate. The accuracy of
these calculations is entirely dependent on the trial wave function,
however, and so the development of accurate wave functions is vital both
in the accurate estimation of physical properties and in understanding 
how certain physical phenomena may be simply represented
in the wave function. 

A widely used trial correlated wave function is the 
Jastrow-Slater\ \cite{FahyWang} or Feenberg\ \cite{Feenberg} form, 
\begin{eqnarray}
  \Psi({\bf r}_1,\dots,{\bf r}_N) =
  \exp\left[-\sum_{i<j} u({\bf r}_i,{\bf r}_j) 
  + \sum_{i} \chi({\bf r}_i) \right] D~.
  \label{eq:JastrowPsi}
\end{eqnarray}
Here $D$ is a Slater determinant of single-particle orbitals
and interparticle correlation is introduced with the two-body term
$u$ in the Jastrow factor. 
The one-body term $\chi$ could in principle be absorbed into the 
single-particle orbitals of the determinant, but it may be 
convenient to retain it explicitly in the Jastrow factor.
In practice, the orbitals in the Slater determinant are often 
determined from a HF or DFT calculation.\ \cite{FahyWang}

In this paper, we will focus on the optimization of Jastrow-Slater 
wave functions in the context of the electronic structure of 
periodic solids. We will apply wave function optimization to 
examine some consequences which emerge from a complete treatment of
inhomogeneity in the two-body correlation term for diamond and 
rhombohedral graphite.

The general form of wave function in Eq.\ \ref{eq:JastrowPsi} has 
been used as the starting point of several methods, including Fermi 
hypernetted chain\ \cite{FHNC} (FHNC) and 
VMC\ \cite{FahyWang} calculations. 
The traditional approach is to use a variational principle on the energy 
(or, in the ``variance minimization'' method,\ \cite{varmin} 
on the fluctuations of the 
energy) to define a best approximation to the true eigenstate within the 
variational freedom allowed by the wave function ansatz.
Expectation values of the energy and other quantities are calculated 
from the trial wave function, approximately in the FHNC approach, 
exactly in VMC (within statistical error of the sampling).
While the VMC method has been used in calculations of periodic solids, 
in practice most calculations have included only homogeneous two-body 
terms in the Jastrow factor\ \cite{FahyWang} 
and the optimization of wavefunctions with
very large numbers of parameters has remained problematical.
To our knowledge, the FHNC approach has not been applied in fully 
three-dimensional electronic structure calculations.

Wave functions determined by the VMC approach are often used as 
guiding, trial functions for the diffusion Monte Carlo (DMC) method.\ \cite{CepAldDMC}
In the DMC context, the accuracy of the wavefunction affects the
numerical efficiency of energy calculations and the accuracy of
other physical quantities. 
When DMC is used in conjunction with 
non-local pseudopotentials,\ \cite{DMCnlpot}
as is commonly the case in chemical applications, an accurate
trial wavefunction is essential for an accurate calculation of the
ground state energy.

We present here, for the first time, a numerically robust, rapidly
convergent iterative method which minimizes the variational energy with 
respect to a very general inhomogeneous form of the Jastrow factor,
including all one- and two-body terms compatible with crystal symmetry.
The method remains numerically well-conditioned, even for large systems
(of the order of several hundred electrons), where there are more than
3000 independent variational parameters in the Jastrow factor.
Within acceptable computational demands on the Monte Carlo sampling, 
the optimal values of parameters in the wave function are found to be
well determined, even when their contribution to the total energy is 
extremely small.

The method places no restrictions on the functional form of the 
anti-symmetric (determinantal) part of the many-body wave function
and can be used in conjunction with related methods recently developed
for energy minimization with respect to all orbitals in the 
determinant\ \cite{1bodyEFO}
or with respect to configuration weights in a multi-determinant 
function.\ \cite{Schautz,Nightingale} 
Taken in conjunction with these methods, the approach presented here
completes the solution to the problem of energy minimization with respect 
to the most general variational terms in wavefunctions of the 
Jastrow-Slater type, as well as in wave functions where the
Jastrow factor multiplies a multi-determinant function.
Although many specific details of the work here refer to periodic
systems, the basic method for Jastrow factor optimization could also be 
used in similar calculations of molecular, atomic, or nuclear structure.

Other general methods exist to achieve energy minimization with 
respect to parameters in many-body wave functions. 
When only one or two parameters are optimized, it is possible
to perform a systematic search of parameter space, as McMillan
did in his pioneering VMC study of the properties of liquid He.\ \cite{McMillan}
The stochastic gradient approximation (SGA) used by 
Harju {\it et al.}\ \cite{SGAHarju} applies control theory to determine 
iterative corrections to the wave function parameters. 
Lin {\it et al.}\ \cite{LinZhang} explicitly computed analytical derivatives 
of the energy with respect to variational parameters and used these to 
optimize their wave function with a Newton-style method. 
However, none of these methods have as yet been applied in systems 
where a very large number of parameters are to be optimized.

In recent years, the variance minimization method of Umrigar 
{\it et al.},\ \cite{varmin} which optimizes the wave function by 
reducing the magnitude of the variance of the local energy, has been 
used much more widely than energy minimization in determining optimal 
parameters for variational wave functions. 
Although not strictly equivalent to energy minimization for a non-exact 
wave function, in practice this method has been very successful in 
improving total energies, providing extremely accurate wave functions
in certain atomic systems.\ \cite{varmin} 
Unfortunately, variance minimization is often subject to confinement to local 
minima and requires much human interaction and experience for successful 
implementation when large numbers of parameters are to be optimized.

Variance minimization may be thought of as fitting energy fluctuations
to a constant, and attempting to reduce the cost function involved in this
fit by direct variation of the wave function parameters. 
The standard approach is to use a least-squares fit of 
these energy fluctuations.\ \footnote{
  This may leave variance minimization susceptible to instability 
  due to the large cost associated with outliers 
  (energies far from the constant value used in the fit).
  Recently, robust estimation methods have been discussed 
  to improve the convergence 
  and stability of variance minimization by fitting energy fluctuations, 
  assuming that they are not necessarily normally 
  distributed.\ \cite{Bressanini} 
}

Our method of energy minimization 
involves the fitting of energy fluctuations to a given
(non-constant) functional form, which is a linear combination of 
operators associated with variations of the wave function parameters
(see Sec.\ \ref{Sec.EnerMin}).
This fitting allows us to determine a ``predictor'',
which links changes in the wave function to changes in the
fitted energy fluctuations. 
This predictor iteratively guides our method to a self-consistent solution,
where the fitted energy fluctuations are zero.
The derivatives of
the true many-body energy with respect to all
the parameters in the wave function are then also zero 
(within statistical sampling error)
for this final solution.

Our predictor is closely related to the random phase approximation (RPA),
introduced by Bohm and Pines\ \cite{BohmPines} and recently discussed in
the context of inhomogeneous systems.\ \cite{Bevan,Gaudoin}
However, as long as the predictor is sufficiently accurate to
ensure stable convergence of the iterations to the self-consistent solution,
its exact form does not affect the final solution. 
Our method allows us to surpass the approximations of the RPA
and produce explicit trial wave functions of unprecedented accuracy 
for electrons in periodic solids.

We will use the optimized wave functions to study the effects of 
charge inhomogeneity on the correlation factors in diamond, as a 
prototype of strongly bonded insulating systems, and rhombohedral graphite, 
as a prototype of highly anisotropic, inhomogeneous solids.
Although earlier studies\ \cite{FahyWang,Malatesta} 
of these and related systems would indicate 
that homogeneous two-body correlation factors gain a large fraction
of the correlation energy in solids and that inhomogeneous correlation 
factors are unlikely to lower the total variational energy greatly, 
substantial differences in correlation factors often give rise to 
relatively little change in energy.
This is particularly true when one is interested in long-range
correlation, which is energetically very delicate.

The rest of this paper is organized as follows:
In Sec.\ \ref{Sec.wavefn}, 
we present the detailed numerical form of variational 
wave functions to be used in these calculations.
The approach of fitting energy fluctuations
and its use in guiding the iterative solution of the energy
minimization problem is discussed in Sec.\ \ref{Sec.EnerMin}.
In Sec.\ \ref{Sec.Results}, 
we present some results on the application of the
method to the periodic solids, diamond and rhombohedral graphite,
and examine the effects of charge density inhomogeneity on the correlation
factors of these systems.
In Sec.\ \ref{Sec.Discussion} we discuss the results and some computational 
details of the method, illustrating the justification for certain aspects 
of our approach with some tests. 
Finally, in Sec.\ \ref{Sec.Conclusion}, we present the overall conclusions of
this study.

% ------------------------------------------------------------------------
\section{Form of the wave function}
\label{Sec.wavefn}
% ------------------------------------------------------------------------

The Jastrow two-body correlation factor $u({\bf r},{\bf r'})$
in Eq.\ \ref{eq:JastrowPsi} can in principle always be expressed
as:
\begin{eqnarray}
  u({\bf r},{\bf r'}) = \sum_{\alpha \beta} 
   f_{\alpha}({\bf r})^* u_{\alpha \beta} f_{\beta}({\bf r'}) ~,
  \label{eq:2bodyexpansion}
\end{eqnarray}
where $f_{\alpha}$ form a complete set of functions and 
$u_{\alpha \beta}$ are expansion coefficients.
(We use $^*$ to indicate complex conjugation throughout this paper).
Similarly, the one-body function $\chi ({\bf r})$ may also be expanded
in a basis set of complete functions $g_{\gamma}$, as:
\begin{eqnarray}
  \chi({\bf r}) = \sum_{\gamma} \chi_{\gamma} g_{\gamma}({\bf r}) ~ .
  \label{eq:1bodyexpansion}
\end{eqnarray}

Apart from an additional term to handle the 
electron-electron cusp,\ \cite{KatoCusp}
we will express the Jastrow factor in the general form given
by Eqs.\ \ref{eq:2bodyexpansion} and\ \ref{eq:1bodyexpansion}.
Values of the parameters $\{ u_{\alpha \beta}, \chi_{\gamma} \}$
will then be determined so that the energy of the system is stationary
with respect to all variations $\{ du_{\alpha \beta}, d\chi_{\gamma} \}$.

% ------------------------------------------------------------------------
\subsection{The electron-electron cusp}
\label{Sec.ElecCusp}
% ------------------------------------------------------------------------

Due to the divergence of the Coulomb interaction between 
two electrons, the correct two-body correlation term 
$u({\bf r},{\bf r'})$ has a cusp where 
${\bf r} \rightarrow {\bf r'}$, leading to slow 
convergence of any expansion in smooth functions of the form in 
Eq.\ \ref{eq:2bodyexpansion}.\ \cite{Morgan} 
For this reason, it is numerically convenient to re-write the
two-body function $u$ in the form:
\begin{eqnarray}
  u({\bf r},{\bf r'}) + u_{\rm sr}(|{\bf r}-{\bf r'}|),
  \label{eq:shortlongu}
\end{eqnarray}
where $u_{\rm sr}(r)$ is a short-ranged (homogeneous) function which has the
correct electron-electron cusp as $r\rightarrow 0$ and 
$u({\bf r},{\bf r'})$ is now a smooth, cuspless function.
We use a form of the short-ranged function $u_{\rm sr}$ which
is generated from a numerical solution of the electron-electron
scattering problem. A discussion of the
generation of $u_{\rm sr}$ is provided in Appendices A and C of 
Ref.~\onlinecite{Cusppaper}.
We define $u_{\rm sr} \equiv - \ln J_{\rm sr}$, where $J_{\rm sr}$ is defined
within Appendix C of Ref.~\onlinecite{Cusppaper}. 
The expansion of the remaining function $u({\bf r},{\bf r'})$
in the form given in Eq.\ \ref{eq:2bodyexpansion} 
then converges much more rapidly
than that of the original function,
for any set of smooth functions $f_{\alpha}$.

In this paper, we generate the short-range function $u_{\rm sr}$ in
a spin-dependent form, to maintain the cusp conditions.\ \cite{KatoCusp}
The cuspless function $u({\bf r},{\bf r'})$ used in our work
is independent of electron spin, although we expect that a spin-dependent
form is as easily optimized. 

% ------------------------------------------------------------------------
\subsection{Separation of one- and two-body terms}
\label{Sec.Sep12}
% ------------------------------------------------------------------------

Summing $u({\bf r},{\bf r'})$ over all electron pairs 
in the basis $f_{\alpha}$ leads to
\begin{eqnarray}
  \sum_{i<j}u({\bf r}_i, {\bf r}_j) & = &
       \frac{1}{2} \sum_{\alpha \beta} u_{\alpha \beta}
       \sum_{i \neq j} f_{\alpha}({\bf r}_i)^* f_{\beta}({\bf r}_j) ~.
  \label{eq:2bodycorrelation}
\end{eqnarray}
This may be thought of as a two-body expansion of the correlation term.
However, it is important to realize that any separation of 
``one-body'' and ``two-body'' terms in the Jastrow factor is somewhat
arbitrary. Removal of terms where $i = j$, as in 
Eq.\ \ref{eq:2bodycorrelation}, is not sufficient to decouple
one- and two-body terms completely. To see this, consider the transformation
of each of the basis functions obtained by subtracting a constant,
$f_{\alpha}'({\bf r}) = f_{\alpha}({\bf r}) - c_{\alpha}$. 
The set remains complete and the function
\begin{eqnarray}
  u({\bf r},{\bf r'}) = \sum_{\alpha \beta} 
   [f_{\alpha}({\bf r})^* - c_{\alpha}^*] u_{\alpha \beta} 
   [f_{\beta}({\bf r'}) - c_{\beta}]  ~,
  \label{eq:arbconstexpansion}
\end{eqnarray}
may be interpreted as a ``two-body'' function in the basis set $f_{\alpha}'$.
However, expanding this correlation factor over all electron pairs,
we find that, in terms of the original basis $f_{\alpha}$,
an additional one-body contribution appears:
\begin{eqnarray}
  \lefteqn{ \sum_{i<j}u({\bf r}_i, {\bf r}_j) =
       \frac{1}{2} \sum_{\alpha \beta} u_{\alpha \beta}
       \sum_{i \neq j} f_{\alpha}({\bf r}_i)^* f_{\beta}({\bf r}_j) } 
       \hspace{0.5cm} \nonumber \\
  & - & \frac{N-1}{2} \sum_{\alpha \beta} \sum_i 
        [u_{\alpha \beta} f_{\alpha}({\bf r}_i)^* c_{\beta} 
      +  u_{\alpha \beta} c_{\alpha}^* f_{\beta}({\bf r}_i)] 
      \label{eq:12bodyseparation}  \\
  & + & ~{\rm constant} ~. \nonumber
\end{eqnarray}
We may regard this as a transformation of the one-body function in
Eq.\ \ref{eq:JastrowPsi}: 
$\chi({\bf r}) \rightarrow \chi({\bf r}) + \chi^0({\bf r})$, 
where the additional term comes from the second line of
Eq.\ \ref{eq:12bodyseparation}: 
\begin{eqnarray}
  \label{eq:natural1body}
  \lefteqn{\chi^0({\bf r}) = \frac{N-1}{2} \times} 
  \hspace{0.5cm} \\ 
  & & \sum_{\alpha} \left\{
        \left[ \sum_{\beta} u_{\alpha \beta} c_{\beta} \right] 
        f_{\alpha}({\bf r})^* 
     +  \left[ \sum_{\beta} u_{\beta \alpha} c_{\beta}^* \right]
        f_{\alpha}({\bf r}) \right\} \nonumber 
\end{eqnarray}

To make a definite numerical separation of our one- and two-body
expansions, we need to consider an appropriate choice of the arbitrary
constants $c_{\alpha}$, noting that $u({\bf r}, {\bf r'})$ is intended 
primarily to affect correlation properties 
(i.e., two-body properties) of the system, leaving single-particle
properties unchanged.
For example, the mean-field
methods that produce $D$ 
[HF, DFT under the local density approximation\ \cite{KohnSham} (LDA), etc.] 
normally give very accurate
single-particle densities, which can be altered substantially by
inclusion of an arbitrary function $u({\bf r},{\bf r'})$,
in $\Psi$ (Eq.\ \ref{eq:JastrowPsi}).
Any substantial change in the density from the HF solution is
likely to be energetically very costly, so that ideally we
would like to decouple changes in $u({\bf r},{\bf r'})$ from
changes in the density. Necessary changes in the density may be
allowed for by optimization of the explicit one-body term $\chi$ in the
Jastrow factor (Eq.\ \ref{eq:1bodyexpansion}) or by the methods of
Ref.~\onlinecite{1bodyEFO}.

Therefore, we would like to choose the constants $c_{\alpha}$ 
such that the average of any one-body operator (such as the density) 
for $\Psi$ remains stationary with respect to variations in the
coefficients $u_{\alpha \beta}$, at least in the absence of 
interparticle correlation.
If we define the one-body operator 
\begin{eqnarray*}
  \rho_{\gamma}({\bf R}) \equiv \sum_i f_{\gamma}({\bf r_i})^* ~ ,
\end{eqnarray*}
for the many-body configuration ${\bf R} = ({\bf r}_1 \dots {\bf r}_N)$, 
then its expectation value is
\begin{eqnarray*}
  \langle \rho_{\gamma} \rangle =
  \langle \Psi | \rho_{\gamma} | \Psi \rangle =
  \int \! f_{\gamma}({\bf r})^* \rho({\bf r}) \ d{\bf r} ~,
\end{eqnarray*}
where $\rho({\bf r})$ is the single particle density.
Since the $f_{\gamma}$ are fixed functions, variations in the
expectation values $\langle \rho_{\gamma} \rangle$ correspond
to variations in $\rho({\bf r})$.
The derivative of $\langle \rho_{\gamma} \rangle$ with respect to 
variations of the $u_{\alpha \beta}$ in Eq.\ \ref{eq:arbconstexpansion} is
(see Appendix\ \ref{App.deriv}) 
\begin{eqnarray*}
  \frac{1}{2} \frac{\partial \langle \rho_{\gamma} \rangle}
                   {\partial u_{\alpha \beta}} 
  = \langle  
    [\rho_{\gamma}({\bf R}) - \langle \rho_{\gamma} \rangle]
    \sum_{i \neq j}[f_{\alpha}({\bf r}_i)^* - c_{\alpha}^*] 
                   [f_{\beta}({\bf r}_j) - c_{\beta}]
    \rangle \\
  = \sum_k \sum_{i \neq j} \langle 
    [f_{\gamma}({\bf r}_k)^* - \langle f_{\gamma} \rangle^*]
    [f_{\alpha}({\bf r}_i)^* - c_{\alpha}^*]
    [f_{\beta}({\bf r}_j) - c_{\beta}]
    \rangle ~ .
\end{eqnarray*}
 
In the absence of interparticle correlation,
\begin{eqnarray*} 
  \langle 
    f_{\gamma}({\bf r}_k)^* f_{\alpha}({\bf r}_i)^* f_{\beta}({\bf r}_j)
  \rangle =
  \left\{
  \begin{array}{ll}
    \langle f_{\gamma}^* f_{\alpha}^* \rangle
    \langle f_{\beta} \rangle
    & \mbox{if $k=i$} \\
    \langle f_{\gamma}^* f_{\beta} \rangle
    \langle f_{\alpha} \rangle^*
    & \mbox{if $k=j$} \\
    \langle f_{\gamma} \rangle^*
    \langle f_{\alpha} \rangle^*
    \langle f_{\beta} \rangle
    & \mbox{otherwise} 
  \end{array}
  \right.
\end{eqnarray*}
The second summation excludes $i = j$, so averages of triple products
never arise.
Thus, in the absence of correlation, the derivative becomes
\begin{eqnarray}
  \lefteqn{\frac{1}{2} \frac{\partial \langle \rho_{\gamma} \rangle}
                {\partial u_{\alpha \beta}} \approx} \hspace{0.5cm}
  \label{eq:approxderiv} \\
  & \sum_{i \neq j} &
       \langle [ f_{\gamma}({\bf r}_i)^* - \langle f_{\gamma} \rangle^* ]
               [ f_{\alpha}({\bf r}_i)^* - c_{\alpha}^* ] \rangle
       \langle   f_{\beta}({\bf r}_j) - c_{\beta}  \rangle
  \nonumber \\
  & + & 
       \langle [ f_{\gamma}({\bf r}_j)^* - \langle f_{\gamma} \rangle^* ]
               [ f_{\beta}({\bf r}_j) - c_{\beta} ] \rangle
       \langle   f_{\alpha}({\bf r}_i)^* - c_{\alpha}^*  \rangle ~.
  \nonumber
\end{eqnarray}
We can guarantee that the right hand side of Eq.\ \ref{eq:approxderiv}
is zero if $c_{\alpha} = \langle f_{\alpha} \rangle$, for all $\alpha$. 
In other words, one-body expectation values remain approximately unaffected by 
the presence of the correlation factor $u({\bf r}, {\bf r'})$, 
provided we expand 
$u({\bf r}, {\bf r'})$ in a basis of ``fluctuation functions'',
$f_{\alpha}'({\bf r}) = f_{\alpha}({\bf r}) - \langle f_{\alpha} \rangle$.
Equivalently, we may retain the original basis $f_{\alpha}$ and form
a one-body term $\chi^0$ from Eq.\ \ref{eq:natural1body}
with $c_{\alpha} = \langle f_{\alpha} \rangle$.
This may then be inserted into the wave function of Eq.\ \ref{eq:JastrowPsi}
and varied as the parameters $u_{\alpha \beta}$ are varied.

Correlation effects are of course present in the actual wave function
used. However, we find that Eq.\ \ref{eq:approxderiv} remains
approximately true, as previously observed.\ \cite{Malatesta,FahyNATO}
For the energy minimization problem, we find that mixed derivatives
of the energy 
$\partial^2 E / \partial u_{\alpha \beta} \partial \chi_{\gamma}$
are approximately zero when $c_{\alpha} = \langle f_{\alpha} \rangle$.
This gives the numerical advantage that minimization of the
energy with respect to $u_{\alpha \beta}$ approximately decouples
from minimization with respect to $\chi_{\gamma}$.
We note that this approximate decoupling holds, even if the relation
$c_{\alpha} = \langle f_{\alpha} \rangle$ is not exactly true.
Thus we may use $c_{\alpha} = \langle D | f_{\alpha} | D \rangle$,
(i.e., LDA or HF averages of $f_{\alpha}$), in place of 
$\langle \Psi | f_{\alpha} | \Psi \rangle$, while still
maintaining the numerical advantages of approximately satisfying
$\partial^2 E / \partial u_{\alpha \beta} \partial \chi_{\gamma} = 0$.

% ------------------------------------------------------------------------
\subsection{Fourier Expansion}
\label{Sec.FourierExpansion}
% ------------------------------------------------------------------------

In the context of periodic systems, it is natural to expand the correlation
function $u({\bf r}, {\bf r'})$ as a Fourier series, where the 
basis functions are
$f_{\bf q} = \exp[i{\bf q \cdot r}]$, for each wave vector ${\bf q}$.
We note that 
\begin{eqnarray}
  \rho_{\bf q}({\bf R}) = \sum_i \exp[-i{\bf q \cdot r}_i]
                        = \sum_i f_{\bf q}({\bf r}_i)^* 
  \label{eq:fouriercharge}
\end{eqnarray}
is the Fourier coefficient of the instantaneous charge density,
given the electron configuration ${\bf R}$.
(In atomic units the electronic charge $e=1$). 
Also, summing over pairs leads to a quadratic product of Fourier 
coefficients, with some modification to remove terms with $i=j$:
\begin{eqnarray}
  \sum_{i < j} f_{\bf q}({\bf r}_i)^* f_{\bf q'}({\bf r}_j)
  & \equiv & \frac{1}{2} ( \rho_{\bf q} \rho_{\bf q'}^* )_{[i \neq j]}.
  \label{eq:1bodyremoved}
\end{eqnarray}

In order to approximately remove the effect of
the two-body terms on the single-particle density, 
following Sec.\ \ref{Sec.Sep12}, we 
subtract appropriate constants from each basis function to
produce a new basis, 
the collective ``charge fluctuation'' coordinates,
\begin{eqnarray}
  \sum_i f'_{\bf q}({\bf r}_i)^* \equiv 
  \Delta \! \rho_{\bf q} = \rho_{\bf q} - \langle \rho_{\bf q} \rangle
                         = \sum_i f_{\bf q}^*({\bf r}_i) - 
                          \langle f_{\bf q} \rangle^* ~ .
  \label{eq:deltarho}
\end{eqnarray}
These provide a suitable expansion of the 
two-body correlation factor that approximately preserves 
single-particle densities. 

A correlation factor using such coordinates was first suggested by
Bohm and Pines\ \cite{BohmPines} for the homogeneous electron gas,
and has recently been discussed by Malatesta {\it et al.}\ \cite{Malatesta} and 
Gaudoin {\it et al.}\ \cite{Gaudoin} in the context of inhomogeneous systems.
In homogeneous systems the expectation value 
$\langle \rho_{\bf q} \rangle$ of each charge
density Fourier coefficient is zero, and so charge density
fluctuations are simply $\rho_{\bf q}$.
For inhomogeneous systems, in general 
$\langle \rho_{\bf q} \rangle \neq 0$ when ${\bf q = G}$, 
a reciprocal lattice vector.

The alternative to using fluctuation coordinates
$\Delta \! \rho_{\bf q}$ is to incorporate 
the equivalent one-body term in the Jastrow factor, 
which in periodic systems is of the form
\begin{eqnarray*}
  \sum_i \chi^0({\bf r}_i) = \sum_{\bf G} \chi^0_{\bf G} \rho_{\bf G}^* ~,
\end{eqnarray*}
with the coefficients coming from the two-body term
\begin{eqnarray}
  \chi^0_{\bf G} = \frac{N-1}{N} \sum_{\bf G'} u_{\bf G G'}
                 \langle \rho_{\bf G'} \rangle ~ , 
  \label{eq:1bodycoeffs}
\end{eqnarray}
as discussed by Malatesta {\it et al.}\ \cite{Malatesta} and
Gaudoin {\it et al.}\ \cite{Gaudoin}

The properties of the Fourier basis $f_{\bf q}$ and the 
correlation factor $u$ lead to some convenient 
symmetry properties for the coefficients $u_{\bf q q'}$. 
The complex conjugate $u_{\bf q q'}^* = u_{\bf -q \, -q'}$,
just as $f_{\bf q} = f_{\bf -q}^*$. 
The exchange symmetry of u, i.e., 
$u({\bf r}, {\bf r'}) = u({\bf r'}, {\bf r})$
implies that $u_{\bf q q'} = u_{\bf -q' \, -q}$.
And, if u possesses inversion symmetry, i.e. 
$u({\bf r}, {\bf r'}) = u({\bf -r}, {\bf -r'})$,
then each $u_{\bf q q'}$ is a real number.
In periodic systems,
\begin{eqnarray*}
  u({\bf r+L}, {\bf r'+L}) = u({\bf r}, {\bf r'})
\end{eqnarray*}
for any Bravais lattice vector ${\bf L}$.
This implies that all Fourier coefficients $u_{\bf q q'}$ are zero
unless ${\bf q - q' = G}$, a reciprocal lattice vector.
Thus, translation symmetry greatly reduces the number of variational
parameters in the two-body terms.

We arrange the wave function in the form 
\begin{eqnarray*}
  \Psi = J_{\rm sr} J_{\rm ih} J_{\rm 1b} D ,
\end{eqnarray*}
isolating the short range component of the Jastrow factor as
\begin{eqnarray*}
  J_{\rm sr} = \exp \left[ \sum_{|{\bf G}| < G_c}
               \chi^0_{\rm sr}({\bf G}) \rho_{\bf G}^*
             - \sum_{i<j} u_{\rm sr}(r_{ij}) \right] ~.
\end{eqnarray*}
The one-body term here is derived from the short range
correlation factor $u_{\rm sr}$ of Eq.\ \ref{eq:shortlongu}, as 
$\chi^0_{\rm sr}({\bf G}) = 
 \tilde{u}_{\rm sr}({\bf G}) \langle \rho_{\bf G} \rangle$,
where $\tilde{u}_{\rm sr}({\bf G})$ is the Fourier transform of
$u_{\rm sr}$ for the reciprocal lattice vector ${\bf G}$.
The prefactor of $(N-1)/N$, which should be present from 
Eq.\ \ref{eq:1bodycoeffs}, approaches unity for
large systems and so is neglected.
For computational efficiency we leave this one-body term in its
Fourier space representation and
$G_c$ is a cut-off chosen for the Fourier sum such that it is
converged within a required accuracy. 

The remaining inhomogeneous part of the two-body Jastrow factor 
is expanded using charge density fluctuation coordinates:
\begin{eqnarray}
  \label{eq:inhomJast}
  J_{\rm ih} = \exp \left[ - \sum_{\bf q} \sum_{\bf G G'}
    u_{\bf q+G \ q+G'} P_{\bf q+G \ q+G'}
    \right]~,
\end{eqnarray}
where
$P_{\bf q+G \ q+G'} \equiv
 (\Delta \rho_{\bf q+G} \Delta \rho_{\bf q+G'})_{[i \neq j]}$,
using the notation $(\cdot)_{[i \neq j]}$, as defined in 
Eq.\ \ref{eq:1bodyremoved} and the definition of $\Delta \rho_{\bf q+G}$ in
Eq.\ \ref{eq:deltarho}.
In practice, this double sum is truncated by using vectors
${\bf q+G}$ of magnitude less than a suitably chosen cut-off $k_c$.

We also allow for one-body optimization through the use
of the explicit one-body Jastrow factor. This one-body Jastrow
factor is also expanded in fluctuation coordinates,
\begin{eqnarray}
  \label{eq:1bJast}
  J_{\rm 1b} = \exp \left[ \sum_{|{\bf G}| < G_c}
               \chi_{\bf G} \Delta \! \rho_{\bf G}^* \right],
\end{eqnarray}
since including the constant 
average values $\langle \rho_{\bf G} \rangle$ merely
adjusts the normalization of the wave function.

The coefficients $u_{\bf q+G \ q+G'}$ and $\chi_{\bf G}$,
defined in Eqs.\ \ref{eq:inhomJast} and\ \ref{eq:1bJast},  
are the final variational parameters of our wave function.
The remaining sections of this paper describe the method
we use to optimize these parameters such that the total energy of a
particular electronic system is stationary. A typical calculation
presented below involves the simultaneous optimization of
over 3000 parameters.

% ------------------------------------------------------------------------
\section{Energy Minimization}
\label{Sec.EnerMin}
% ------------------------------------------------------------------------

We wish to optimize the wave function
\begin{eqnarray*}
  \Psi = \Psi( \mbox{\boldmath $\alpha$} )~,
\end{eqnarray*}
where $\mbox{\boldmath $\alpha$}=\{ \alpha_m \}$ a vector of parameters,
by solving the Euler-Lagrange equations,
\begin{eqnarray}
  \label{eq:ELeqns}
  \frac{\partial \langle {\cal H} \rangle}{\partial \alpha_m} = 0
  ~\mbox{for all $m$}~.
\end{eqnarray}
We note that the Hamiltonian for the system is
\begin{eqnarray}
  {\cal H} = - \frac{1}{2} \sum_i \nabla_i^2
                 + \sum_i V_{\rm ext}({\bf r}_i)
                 + \sum_{i<j} V(r_{ij})~,
  \label{eq:Hamiltonian}
\end{eqnarray}
where each sum is over the electrons in the system.

Solving Eq.\ \ref{eq:ELeqns} is equivalent (see Appendix\ \ref{App.deriv})
to solving the following system of equations
\begin{eqnarray}
  \label{eq:ELsystem}
  \langle \Delta \! {\cal H} \ \Delta \! {\cal O}_m \rangle = 0
  ~\mbox{for all $m$}~,
\end{eqnarray}
where,
given a many-body configuration ${\bf R} = \{ {\bf r}_i \}$,
we define: 
$\Delta A({\bf R}) \equiv A({\bf R}) - \langle A \rangle$,
for any operator $A$; 
and the local values of the operators
${\cal O}_m$ as, 
\begin{eqnarray}
  \label{eq:Operdefn}
  {\cal O}_m({\bf R}) &\equiv& 
    \frac{\partial}{\partial \alpha_m} \ln \Psi({\bf R})
    \\
    & = & 
    \left\{ \begin{array}{ll}
      -P_{\bf q q'}({\bf R}) &
      ~\mbox{for $\alpha_m = u_{\bf q q'}$}~,
      \\
      \Delta \rho_{\bf G}({\bf R})^* &
      ~\mbox{for $\alpha_m = \chi_{\bf G}$}~.
      \nonumber
    \end{array} \right.
\end{eqnarray}
We shall refer to the local value of the Hamiltonian operator as
the ``local energy'',
\begin{eqnarray}
  E({\bf R}) &\equiv& \frac{{\cal H} \Psi({\bf R})}{\Psi({\bf R})}~.
  \label{eq:localener}
\end{eqnarray}

We approach the problem of solving the Euler-Lagrange equations
(Eq.\ \ref{eq:ELeqns}) indirectly, by considering systematic fluctuations
of the energy for a given trial wave function $\Psi(\mbox{\boldmath $\alpha$})$.

% ------------------------------------------------------------------------
\subsection{Systematic Energy Fluctuations}
\label{Sec.SysEnerFluct}
% ------------------------------------------------------------------------

Consider fitting the local energy $E({\bf R})$ to the functional form
\begin{eqnarray}
  E_0 + \sum_m V_m {\cal O}_m({\bf R}) ~,
  \label{eq:enerexpansion}
\end{eqnarray}
in the least-squares sense,
where $\{ {\cal O}_m \}$ 
is the set of functions with which we fit the energy, and
$\{ V_m \}$
is the vector of fitting coefficients. 
The least-squares problem reduces to minimizing the integral
\begin{eqnarray*}
  \langle \Psi | \left\{ {\cal H} - E_0
  - \sum_m V_m {\cal O}_m
  \right\}^2 | \Psi \rangle~,
\end{eqnarray*}
which is equivalent (see Appendix\ \ref{App.leastsq}) 
to solving the linear system
\begin{eqnarray}
  \sum_m V_m \langle \Delta {\cal O}_m \Delta {\cal O}_n \rangle = 
  \langle \Delta \! E \Delta {\cal O}_n \rangle 
  ~ \mbox{for all $n$.}
  \label{eq:lstsqsys}
\end{eqnarray}

We recognize immediately that if the functions ${\cal O}_m$ are those
functions associated with variations 
of the wave function parameters $\alpha_m$ (Eq.\ \ref{eq:Operdefn}), 
then the right-hand side of Eq.\ \ref{eq:lstsqsys}
{\it is} the vector of Euler-Lagrange derivatives in Eq.\ \ref{eq:ELsystem}.
Therefore, the Euler-Lagrange equations (Eq.\ \ref{eq:ELeqns}) are solved if 
all the fitted coefficients $V_m$ are zero.\footnote{ 
	If the matrix $\langle \Delta {\cal O}_m \Delta {\cal O}_n \rangle$ 
	is singular, or at least numerically so (i.e., has a very small
        eigenvalue), singular value decomposition is recommended 
        to set the appropriate fitting coefficients $V_l$ to zero. 
        Such problems do not render the wave function parameters $\alpha_l$ 
	ill-conditioned, as the small size of the 
        corresponding expectation values
	$\langle \Delta {\cal O}_l^2 \rangle$ implies that 
        changes in $\alpha_l$ have little effect on the total energy.
}
 
As an illustrative example, consider $\Psi_0$, an eigenstate of ${\cal H}$.
Indeed, the local energy is a constant, independent of 
${\bf R}$, and so we would find that each of the fitted coefficients 
$V_m$ is zero. Therefore, the Euler-Lagrange derivatives
$\langle \Delta {\cal H} \Delta {\cal O}_m \rangle$ are all zero, and so the
energy must be stationary with respect to variations in $\Psi_0$, 
as we would expect for an eigenstate.

For the trial wave function $\Psi(\mbox{\boldmath $\alpha$})$, 
no choice of $\mbox{\boldmath $\alpha$}$ gives an
{\it exact} eigenstate of the Hamiltonian. However, for a particular
choice of parameters, the absence of {\it systematic} variations
of the energy (i.e., variations correlated with the variations of the
functions ${\cal O}_m$) ensures that the fitting coefficients $V_m$
are all zero and that the average energy is stationary with respect to
all variations of the parameters $\mbox{\boldmath $\alpha$}$.
Within the parametric freedom of the trial wave function $\Psi$, this is
our best approximation to an eigenstate.

% ------------------------------------------------------------------------
\subsection{Iterative Procedure}
\label{Sec.IterProc}
% ------------------------------------------------------------------------

We now describe a procedure which aims, by appropriate choice of the
parameters $\mbox{\boldmath $\alpha$}$, to set the fitted coefficients
$V_m$ of the total energy $E$ to zero. 
As defined in Eq.\ \ref{eq:lstsqsys}, these $V_m$ depend on the
wave function $\Psi(\mbox{\boldmath $\alpha$})$ and so are functions 
of the parameters $\mbox{\boldmath $\alpha$}$.
However, the functional dependence
of $V_m$ on $\mbox{\boldmath $\alpha$}$ is not available 
in its exact analytic form and 
we are unable to solve directly the system $V_m(\mbox{\boldmath $\alpha$})=0$, i.e., to
find the root $\mbox{\boldmath $\alpha$}$ which will guarantee the solution to
the corresponding Euler-Lagrange equations.

Instead, using the wave function $\Psi(\mbox{\boldmath $\alpha$}^0)$, 
for a particular choice of the parameters $\mbox{\boldmath $\alpha$}^0$, we construct a
predictor function 
$V'_m(\mbox{\boldmath $\alpha$}; \mbox{\boldmath $\alpha$}^0)$, which
approximates the unknown function $V_m(\mbox{\boldmath $\alpha$})$ for general values
of $\mbox{\boldmath $\alpha$}$. 
More precisely, we construct 
$V'_m(\mbox{\boldmath $\alpha$}; \mbox{\boldmath $\alpha$}^0)$ so that
\begin{eqnarray}
  V'_m(\mbox{\boldmath $\alpha$}^0; \mbox{\boldmath $\alpha$}^0) 
  \equiv V_m(\mbox{\boldmath $\alpha$}^0)
  \label{eq:predictordefn}
\end{eqnarray}
(i.e., $V'_m$ is exact when 
$\mbox{\boldmath $\alpha$}= \mbox{\boldmath $\alpha$}^0$) and, 
\begin{eqnarray}
  V'_m(\mbox{\boldmath $\alpha$}; \mbox{\boldmath $\alpha$}^0) 
  \approx V_m(\mbox{\boldmath $\alpha$})~,
  \label{eq:predictorapprox}
\end{eqnarray}
for all relevant values of $\mbox{\boldmath $\alpha$}$.

To determine this predictor, 
we use the specific form of the Hamiltonian (Eq.\ \ref{eq:Hamiltonian})
and trial wave function $\Psi$ (Sec.\ \ref{Sec.wavefn}), and
we partition the local energy $E({\bf R})$ into a sum of contributions,
\begin{eqnarray}
  E({\bf R}) = \sum_i \epsilon^{(i)}({\bf R})~,
  \label{eq:enercontributions}
\end{eqnarray}
where the $\epsilon^{(i)}$ come from various terms in
the kinetic and potential energy (see below).
Each contribution $\epsilon^{(i)}({\bf R})$ is approximated with
the functional form
\begin{eqnarray}
  \epsilon^{(i)}_0 + \sum_m v^{(i)}_m {\cal O}_m({\bf R})~.
  \label{eq:enerfuncform}
\end{eqnarray}
For some terms, using the specific form of the local energy for 
$\Psi(\mbox{\boldmath $\alpha$})$, 
we can expand $\epsilon^{(i)}$ analytically as in Eq.\ \ref{eq:enerfuncform},
enabling us to determine, exactly or approximately, 
the function $v^{(i)}_m(\mbox{\boldmath $\alpha$})$. In this analytic form, 
$v^{(i)}_m(\mbox{\boldmath $\alpha$})$ is independent of the choice of 
$\mbox{\boldmath $\alpha$}^0$ and
so remains equally valid for all $\mbox{\boldmath $\alpha$}$.

Where analytic expressions are too complex to derive, we may approximate
$v^{(i)}_m(\mbox{\boldmath $\alpha$})$ by fitting $\epsilon^{(i)}$ to Eq.\ \ref{eq:enerfuncform}. 
The fitting coefficients for $\epsilon^{(i)}$, are found 
by solving the analogue of Eq.\ \ref{eq:lstsqsys},
\begin{eqnarray}
  \sum_m v^{(i)}_m
  \langle \Delta {\cal O}_m \Delta {\cal O}_n \rangle = 
  \langle \Delta \epsilon^{(i)} \Delta {\cal O}_n \rangle 
  ~ \mbox{for all $n$,}
  \label{eq:contributionlstsqsys}
\end{eqnarray}
where $v^{(i)}_m = v^{(i)}_m(\mbox{\boldmath $\alpha$}^0)$ is determined by using 
$\Psi(\mbox{\boldmath $\alpha$}^0)$ to evaluate the required expectation values.
This produces the value of each coefficient at $\mbox{\boldmath $\alpha$}^0$.
We may also fit the derivatives of $\epsilon^{(i)}$ with respect
to $\mbox{\boldmath $\alpha$}$, to determine the 
linear dependence of each $v^{(i)}_m(\mbox{\boldmath $\alpha$})$.
We may then approximate the function
\begin{eqnarray}
  \label{eq:Taylorexp}
  \lefteqn{v^{(i)}_m(\mbox{\boldmath $\alpha$}) \approx} \hspace{0.5cm}
  \\
  & & v^{(i)}_m(\mbox{\boldmath $\alpha$}; \mbox{\boldmath $\alpha$}^0) 
  = v^{(i)}_m(\mbox{\boldmath $\alpha$}^0) +
  \sum_l \left. \frac{\partial v^{(i)}_m}{\partial \alpha_l} 
  \right|_{\mbox{\boldmath $\alpha$}^0}
  \! \! ( \alpha_l - \alpha^0_l )~,  
  \nonumber
\end{eqnarray}
where the values of the derivatives are found by fitting
$\partial \epsilon^{(i)} / \partial \alpha_l$ 
to Eq.\ \ref{eq:enerfuncform} using $\Psi(\mbox{\boldmath $\alpha$}^0)$.
In evaluating the term $\partial v^{(i)}_m / \partial \alpha_l$,
we consider only the explicit variation of the term $v^{(i)}_m$
with $\alpha_l$. We do not include the implicit variation due to 
the dependence of the probability distribution 
$|\Psi(\mbox{\boldmath $\alpha$})|^2$
on $\mbox{\boldmath $\mbox{\boldmath $\alpha$}$}$. 
In practice, the only term for which we need to
fit $\partial \epsilon^{(i)} / \partial \alpha_l$
is explicitly linear in the parameters
$\mbox{\boldmath $\mbox{\boldmath $\alpha$}$}$ 
and so the linear expansion in Eq.\ \ref{eq:Taylorexp} is
valid over a wide range of values of 
$\mbox{\boldmath $\alpha$}$.

Just as the energy contributions $\epsilon^{(i)}({\bf R})$ partition the
local energy, we may regard the analytic and fitted coefficients
$v_m^{(i)}$ as an approximate partition of the local energy coefficients
$V_m$. We define this partition as follows:
\begin{eqnarray*}
  W_m(\mbox{\boldmath $\alpha$}; \mbox{\boldmath $\alpha$}^0)
  \equiv
  S_m(\mbox{\boldmath $\alpha$}) 
  + T_m(\mbox{\boldmath $\alpha$}; \mbox{\boldmath $\alpha$}^0)~,
\end{eqnarray*}
where the sum of analytically derived coefficients is
\begin{eqnarray*}
  S_m(\mbox{\boldmath $\alpha$}) \equiv 
    \sum_{\rm analytic} v_m^{(i)}(\mbox{\boldmath $\alpha$})~,
\end{eqnarray*}
and the sum of numerically determined coefficients, evaluated using
$\Psi(\mbox{\boldmath $\alpha$}^0)$ and Eq.\ \ref{eq:contributionlstsqsys}, is
\begin{eqnarray*}
  T_m(\mbox{\boldmath $\alpha$};\mbox{\boldmath $\alpha$}^0) 
  \equiv
  \sum_{\rm fitted} 
    v_m^{(i)}(\mbox{\boldmath $\alpha$};\mbox{\boldmath $\alpha$}^0)~.
\end{eqnarray*}
We construct the predictor 
$V'_m(\mbox{\boldmath $\alpha$}; \mbox{\boldmath $\alpha$}^0)$
such that it satisfies Eq.\ \ref{eq:predictordefn}, i.e., 
\begin{eqnarray}
  V'_m(\mbox{\boldmath $\alpha$}; \mbox{\boldmath $\alpha$}^0) = 
  V_m(\mbox{\boldmath $\alpha$}^0) +
    W_m(\mbox{\boldmath $\alpha$};\mbox{\boldmath $\alpha$}^0)
    - W_m(\mbox{\boldmath $\alpha$}^0;\mbox{\boldmath $\alpha$}^0)
  \label{eq:predictor}
\end{eqnarray}
where $V_m(\mbox{\boldmath $\alpha$}^0)$ are found by 
solving Eq.\ \ref{eq:lstsqsys}. 

We define our iterative approach to determining the parameters 
$\mbox{\boldmath $\alpha$}$,
for which the coefficients $V_m(\mbox{\boldmath $\alpha$})$ 
are zero, as follows:
\vspace{0.5cm}
\begin{enumerate}
  \item Given the set of parameters $\mbox{\boldmath $\alpha$}^n$, 
        construct the wave function
        $\Psi(\mbox{\boldmath $\alpha$}^n)$.
        \label{iter:constrwavefn}
  \item Evaluate the required expectation values 
        in Eqs.\ \ref{eq:lstsqsys} 
        and\ \ref{eq:contributionlstsqsys}, using 
        $\Psi(\mbox{\boldmath $\alpha$}^n)$, to
        find the fitting coefficients $V_m(\mbox{\boldmath $\alpha$}^n)$ and 
        numerical functions 
        $v^{(i)}_m(\mbox{\boldmath $\alpha$};{\mbox{\boldmath $\alpha$}^n})$. 
        \label{iter:fitting}
  \item {\it If} the total energy fitting coefficients are zero, i.e.,
        $V_m(\mbox{\boldmath $\alpha$}^n) = 0 ~\mbox{for all $m$}$, 
        {\it then} we are done, {\it otherwise} continue. 
        \label{iter:donecheck}
  \item Construct the predictor function 
        $V'_m(\mbox{\boldmath $\alpha$};{\mbox{\boldmath $\alpha$}^n})$
        in Eq.\ \ref{eq:predictor} 
        using the fitted terms from Step.\ \ref{iter:fitting}, and
        find the solution $\mbox{\boldmath $\alpha$}^{n+1}$
        to the system
        \begin{eqnarray}
          V'_m(\mbox{\boldmath $\alpha$}^{n+1};{\mbox{\boldmath $\alpha$}^n}) 
          = 0 ~\mbox{for all $m$,}
          \label{eq:generateparam}
        \end{eqnarray}
        using the Newton-Raphson method\ \cite{NumRec} 
        (see Appendix\ \ref{App.NewtonRaphson}).
        Use this set of parameters $\mbox{\boldmath $\alpha$}^{n+1}$ 
        in Step\ \ref{iter:constrwavefn}.
        \label{iter:findroot}
\end{enumerate}
\vspace{0.5cm}

Iterations continue until the total energy coefficients $V_m(\mbox{\boldmath $\alpha$}^n)$
tend to zero, and the values of the parameters $\mbox{\boldmath $\alpha$}^n$ converge.
Note that even though the predictor 
$V'_m(\mbox{\boldmath $\alpha$};{\mbox{\boldmath $\alpha$}^n})$ is only
an approximation to the exact function $V_m(\mbox{\boldmath $\alpha$})$,
Eq.\ \ref{eq:predictordefn} guarantees
that at the converged solution
\begin{eqnarray*}
  0 = V'_m(\mbox{\boldmath $\alpha$};{\mbox{\boldmath $\alpha$}}) = V_m(\mbox{\boldmath $\alpha$})~.
\end{eqnarray*}
In other words, the parameter set $\mbox{\boldmath $\alpha$}$ solves the 
Euler-Lagrange equations for $\langle {\cal H} \rangle$ exactly.
Clearly, the larger the neighbourhood within which the approximate
relation in Eq.\ \ref{eq:predictorapprox} holds, the faster this
iterative procedure will converge. In the trivial case, if the exact analytic
form of $V_m(\mbox{\boldmath $\alpha$})$ were known {\it a priori}, then we could just solve
the Euler-Lagrange equations in one step using a suitable
root-finding method.

Rather than starting the procedure from an initial guess $\mbox{\boldmath $\alpha$} = {\bf 0}$,
we may begin at Step\ \ref{iter:findroot} using only the 
analytic terms in the predictor, since these are independent of the
wave function and so do not require fitting. 
That is, we solve the system
\begin{eqnarray*}
  S_m(\mbox{\boldmath $\alpha$}^1) = 0 ~\mbox{for all}~m~.
\end{eqnarray*} 
The solution set $\mbox{\boldmath $\alpha$}^1$ 
is then used in Step\ \ref{iter:constrwavefn}.

% ------------------------------------------------------------------------
\subsection{Partitioning the local energy}
\label{Sec.EnergyPartition}
% ------------------------------------------------------------------------

We note that the potential energy operators $V_{\rm ext}$ and $V$ 
in the Hamiltonian ${\cal H}$, defined in Eq.\ \ref{eq:Hamiltonian}, are 
multiplicative, and therefore their contributions to 
$E({\bf R})$ are constant with respect to variations of the 
wave function parameters $\mbox{\boldmath $\alpha$}$.
Variations of $\mbox{\boldmath $\alpha$}$ affect only the contributions of the 
differential kinetic energy operator. Thus, solving the Euler-Lagrange
equations for $\langle {\cal H} \rangle$ (Eq.\ \ref{eq:ELeqns}) amounts
to adjusting the systematic fluctuations of the kinetic energy to
cancel those in the potential energy exactly.

If we extract a variational part 
$\phi( \mbox{\boldmath $\alpha$} )$ of $\Psi$,
such that $\Psi = \phi \Psi'$, where $\phi$ is dependent on the set of 
parameters $\mbox{\boldmath $\alpha$}$, and $\Psi'$ is independent of them. 
Then we may partition $E({\bf R})$ as follows: 
\begin{eqnarray*}
  E({\bf R}) = \epsilon^{(1)}({\bf R}) 
             + \epsilon^{(2)}({\bf R}) + \epsilon^{(3)}({\bf R})
\end{eqnarray*}
where we define 
\begin{eqnarray*}
  \epsilon^{(1)}({\bf R}) & \equiv & - \frac{1}{2} \sum_{i=1}^N 
          \frac{\nabla_i^2 \phi({\bf R})}{\phi({\bf R})} ~, 
          \\  
  \epsilon^{(2)}({\bf R}) & \equiv & - \sum_{i=1}^N 
          \frac{\mbox{\boldmath $\nabla$}_i \phi({\bf R})}{\phi({\bf R})}
          {\bf \cdot} \frac{\mbox{\boldmath $\nabla$}_i 
          \Psi'({\bf R})}{\Psi'({\bf R})}
          ~, \\
  \epsilon^{(3)}({\bf R}) & \equiv & - \frac{1}{2} \sum_{i=1}^N 
          \frac{\nabla_i^2 \Psi'({\bf R})}{\Psi'({\bf R})}
          + \sum_{i=1}^N V_{\rm ext}({\bf r}_i)
          + \sum_{i<j}^N V(r_{ij})
          ~ .
\end{eqnarray*}
Clearly, $\epsilon^{(3)}$ is constant with respect to variations 
of $\mbox{\boldmath $\alpha$}$. 
Further analysis of each of $\epsilon^{(1)}$ and $\epsilon^{(2)}$
is necessary to determine how each depends on $\mbox{\boldmath $\alpha$}$. 
However, for $\phi$ expressible in the form of the Jastrow factors 
in Sec.\ \ref{Sec.wavefn}, i.e., 
\begin{eqnarray*}
  \phi(\mbox{\boldmath $\alpha$}) = \exp \left[ 
    \sum_l \alpha_l {\cal O}_{l} \right]~,
\end{eqnarray*}
where ${\cal O}_{l}$ are defined in Eq.\ \ref{eq:Operdefn},
we see that: (i) $\epsilon^{(2)}$
is at most linear in $\mbox{\boldmath $\alpha$}$, but involves terms
coming from $\Psi'$, so that it may be impossible to 
determine analytic expressions for the coefficients
$v_m^{(2)}(\mbox{\boldmath $\alpha$})$;
(ii) $\epsilon^{(1)}$ is at most quadratic in 
$\mbox{\boldmath $\alpha$}$, and involves only $\phi$, for which we have
an analytic expression, and therefore may derive analytic 
approximations to the coefficients $v_m^{(1)}(\mbox{\boldmath $\alpha$})$
(see Sec.\ \ref{Sec.PredictorAnalApprox}).

% ------------------------------------------------------------------------
\subsection{Analytic terms in the predictor}
\label{Sec.PredictorAnalApprox}
% ------------------------------------------------------------------------

The initial predictor $S_m(\mbox{\boldmath $\alpha$})$ used in 
the iterative method involves only analytic terms. 
We determine these by direct expansion of particular
local energy contributions, given the analytic form of the wave function.
We now consider the contributions of the one- and two-body terms
individually.

% ------------------------------------------------------------------------
\subsubsection{One-body Jastrow Factor}
\label{Sec.PredictorAnalApprox.J1b}
% ------------------------------------------------------------------------

Replacing $\phi(\mbox{\boldmath $\alpha$})$ in Sec.\ \ref{Sec.EnergyPartition}
with $J_{\rm 1b}(\mbox{\boldmath $\chi$})$, we derive an analytic
expression for the energy contribution $\epsilon^{(1)}$ 
(see Appendix\ \ref{App.1bEnerCon}).
From this we may extract those coefficients $v_{\bf G}^{(1)}$ 
of the functions ${\cal O}_{\bf G} = \Delta \rho_{\bf G}^*$:
\begin{eqnarray*}
  v_{\bf G}^{(1)}(\mbox{\boldmath $\chi$}) =
    \frac{1}{2} G^2 \chi_{\bf G} +
    \frac{1}{2} \sum_{\bf G'} \chi_{\bf G-G'} ({\bf G - G'}) {\bf \cdot G'}
                                \chi_{\bf G'}~.
\end{eqnarray*}
By assumption, the one-body contribution of $\epsilon^{(3)}$ is zero.
That is, the mean field methods used to calculate the determinant $D$
should remove (approximately) all systematic 
one-body fluctuations in the local energy,
and the use of ``fluctuation functions'' $\Delta \rho_{\bf q}$ in 
the two-body Jastrow factor approximately
removes its effect on one-body operators. 
Therefore, we assume initially that the coefficients $v_{\bf G}^{(3)} = 0$.
We are unable to derive an analytic expression for the coefficients
$v_{\bf G}^{(2)}$ of the energy contribution $\epsilon^{(2)}$ and so,
for the moment, we leave these aside.

Constructing the full analytic approximation 
\begin{eqnarray*}
  S_{\bf G}(\mbox{\boldmath $\chi$}) = 
    v_{\bf G}^{(1)}(\mbox{\boldmath $\chi$})~, 
\end{eqnarray*}
we see that the roots of $S_{\bf G}$ are trivially
$\mbox{\boldmath $\chi$} = {\bf 0}$, as we would expect.

% ------------------------------------------------------------------------
\subsubsection{Two-body Jastrow Factor}
\label{Sec.PredictorAnalApprox.J2b}
% ------------------------------------------------------------------------

According to Sec.\ \ref{Sec.wavefn}, the two-body Jastrow correlation factor
is divided into a short-range term $J_{\rm sr}$  and 
an inhomogeneous term $J_{\rm ih}$.
We optimize the variational parameters ${\bf u}$ in $J_{\rm ih}$.
As for the one-body Jastrow, we expand the corresponding
energy contribution $\epsilon^{(1)}$ which depends on $J_{\rm ih}$ alone
(see Appendix\ \ref{App.2bEnerCon}).
This leads us to an approximation to the coefficients
of the functions ${\cal O}_{\bf q q'} = -P_{\bf q q'}$ in $\epsilon^{(1)}$:
\begin{eqnarray*}
  \lefteqn{v_{\bf q q'}^{(1)}({\bf u}) =} \hspace{0.5cm}
  \\
  && \frac{1}{2} u_{\bf q q'}(q^2 + q'^2)
    + 2 \sum_{\bf k k'} u_{\bf q k} 
                       ({\bf k \cdot k'}) \langle \rho_{\bf k'-k} \rangle 
                       u_{\bf k' q'}
\end{eqnarray*}

Again, we are unable to derive expressions for the coefficients
corresponding to $\epsilon^{(2)}$. We extract a two-body contribution
from the constant contribution $\epsilon^{(3)}$ 
(see Appendix\ \ref{App.2bEnerCon}), as
$v_{\bf q q'}^{(3)} \approx 
 - \frac{1}{2}V({\bf q}) \delta({\bf q' - q})$,
for the electron interaction $V$.
We replace the true interaction in this expression 
with a pseudointeraction $V_{\rm ps}$,
which is generated for a given cut-off radius $r_c$ and reference
eigenvalue $\epsilon$, as explained in Ref.~\onlinecite{Cusppaper}.
$V_{\rm ps}$ is used to generate the short-range
two-body function $u_{\rm sr}$ (Eq.\ \ref{eq:shortlongu}),
used in $J_{\rm sr}$.
The purpose of this modification of $v_{\bf q q'}^{(3)}$ is to
account for the presence of $J_{\rm sr}$ in $\Psi$ and is explained in
Appendix\ \ref{App.ShortJConseq}.

From these contributions we construct the analytic predictor
\begin{eqnarray*}
  S_{\bf q q'}({\bf u}) = 
  v_{\bf q q'}^{(1)}({\bf u}) - 
  \frac{1}{2} V_{\rm ps}({\bf q}) \delta ({\bf q - q'})~.
\end{eqnarray*}
We notice that for periodic systems, this
function is separable in the points ${\bf q}$ of the 
first Brillouin zone (BZ).
For each ${\bf q}$ in BZ, we may expand 
$S_{{\bf q+G},{\bf q+G'}}$ as a function of $\{ u_{{\bf q+G},{\bf q+G'}} \}$,
with no coupling to parameters $u_{{\bf q'+H},{\bf q'+H'}}$ for 
${\bf q' \neq q}$. 
This block diagonal form of the analytic predictor allows us to find the roots
${\bf u}^1$ by solving for each block (i.e., each ${\bf q}$) individually.
We do this using the Newton-Raphson method.

A reliable initial guess for the Newton-Raphson method, rather than using
${\bf u=0}$, is the homogeneous solution of $S_{\bf q q'}=0$.
If we regard the system as homogeneous, i.e., 
$\langle \rho_{\bf G} \rangle = 0$ for ${\bf G \neq 0}$, 
then the solution is
\begin{eqnarray}
  u_{\bf q q'} = \delta({\bf q' - q}) \frac{1}{4N} \left[ 
    \sqrt{ 1 + \frac{8 N V_{\rm ps}({\bf q})}{q^2} } - 1 \right] ~.
  \label{eq:homosoln}
\end{eqnarray}
This $u_{\bf q q'}$ is used as the starting point for the Newton-Raphson
iterations, and for all systems studied, this initial guess produced 
convergent roots of $S_{\bf q q'}=0$.

% ------------------------------------------------------------------------
\subsection{Numerical terms in the predictor}
\label{Sec.PredictorNumer}
% ------------------------------------------------------------------------

To complete the construction of the predictor 
$V'_m(\mbox{\boldmath $\alpha$};\mbox{\boldmath $\alpha$}^0)$
defined in Eq.\ \ref{eq:predictor}, for a given set of 
variational parameters $\mbox{\boldmath $\alpha$}^0$, 
we must determine some terms numerically 
by fitting fluctuations in the energy of the system. 

We calculate $V_m(\mbox{\boldmath $\alpha$}^0)$
in Eq.\ \ref{eq:predictor} by 
fitting the entire local energy, using Eq.\ \ref{eq:lstsqsys}.
Also, we use the fitting method to numerically determine the functions 
$v_m^{(2)}(\mbox{\boldmath $\alpha$};\mbox{\boldmath $\alpha$}^0)$
given in Eq.\ \ref{eq:Taylorexp}, by fitting the energy contribution
$\epsilon^{(2)}$ and each of its derivatives with respect to the parameters
to be optimized.
Again, we note that $\epsilon^{(2)}$ is explicitly linear in the
parameters $\mbox{\boldmath $\alpha$}$ and the approximation in
Eq.\ \ref{eq:Taylorexp} is exact in this case.

To do all this numerical work we use Monte Carlo sampling to determine 
the required expectation values:
\begin{eqnarray*}
  \begin{array}{c}
    \langle \Delta {\cal O}_m \Delta {\cal O}_n \rangle \\
    \langle \Delta E \Delta {\cal O}_m \rangle \\
    \langle \Delta \epsilon^{(2)} \Delta {\cal O}_m \rangle \\
    \langle \Delta ( \frac{\partial \epsilon^{(2)}}{\partial \alpha_m} )
            \Delta {\cal O}_n \rangle
  \end{array}
\end{eqnarray*}
for $m,n$ ranging over the number of parameters $N_{\alpha}$ in the set
$\mbox{\boldmath $\alpha$}$. However, for simultaneous optimization of
one- and two-body terms, we make some simplifications to reduce the
computational workload. 
If we assume that the parameter $\alpha_m$ can be varied
independently of $\alpha_n$, 
this corresponds to assuming that
$\langle \Delta {\cal O}_m \Delta {\cal O}_n \rangle = 0$.

In the expansion of the analytic local energy terms coming from $J_{\rm ih}$,
outlined in Appendix\ \ref{App.2bEnerCon}, we saw that these consisted of
one- and two-body terms. However, the one-body terms contained Fourier
coefficients of the average charge density $\langle \rho_{\bf q} \rangle$,
which are zero for ${\bf q} \neq {\bf G}$, a reciprocal lattice vector.
Also, the analytic form of the predictor (Sec.\ \ref{Sec.PredictorAnalApprox})
is separable in the k-points of the first Brillouin Zone.
Therefore, we make the following approximations to the covariance matrix
$\langle \Delta {\cal O}_m \Delta {\cal O}_n \rangle$:
\begin{eqnarray*}
  \langle \Delta {\cal O}_{{\bf q+G},{\bf q+G'}} 
          \Delta {\cal O}_{{\bf q'+H'},{\bf q'+H'}} \rangle
  &=& 0 ~\mbox{for}~ {\bf q \neq q'}~;
  \\
  \langle \Delta {\cal O}_{{\bf q+G},{\bf q+G'}} \Delta {\cal O}_{\bf H} \rangle
  &=& 0 ~\mbox{for}~ {\bf q \neq 0}~,
\end{eqnarray*}
for reciprocal lattice vectors ${\bf G, G', H, H'}$.
 
We regard one- and two-body optimization as
independent for non-zero k-points in the first Brillouin Zone.
If we also exclude the covariance terms between one- and two-body operators
for ${\bf q = 0}$, we find that this slows the convergence of the method
for smaller systems. 
For larger systems, this exclusion prevents the
convergence of the Newton-Raphson method at the first iteration of our method, 
thus halting the optimization process.
However, we are not, in any way, confined to using the Newton-Raphson method
to find the roots of the predictor,
and other, more robust, root-finding methods 
might overcome this problem. 

Note that $\epsilon^{(2)}$ depends on the variational part $\phi$ of 
the wave function which is being optimized. 
The variational components are
$J_{\rm 1b}$ and $J_{\rm ih}({\bf q})$ for each ${\bf q}$ in BZ,
where we define 
\begin{eqnarray*}
  J_{\rm ih}({\bf q}) = 
  \exp \left[ -\sum_{\bf G G'} u_{{\bf q+G},{\bf q+G'}} 
                              P_{{\bf q+G},{\bf q+G'}} \right]~,
\end{eqnarray*}
so that $J_{\rm ih} = \prod_{\bf q} J_{\rm ih}({\bf q})$.

This greatly reduces the complexity of the predictor,
without sacrificing convergence of the method
for the systems studied here. Ultimately, the predictor
$V'_m(\mbox{\boldmath $\alpha$};\mbox{\boldmath $\alpha$}^0)$ is
itself only an approximation of the true function
$V_m(\mbox{\boldmath $\alpha$})$, but by definition matches this function
at the current values of the parameters being optimized.
Therefore, approximations in the predictor affect only the
rate of convergence of the method. 

The expressions for $\epsilon^{(2)}$ and its derivatives 
$\partial \epsilon^{(2)} / \partial \chi_{\bf G}$ for the one-body
Jastrow factor $J_{\rm 1b}$
are given in Appendix\ \ref{App.1bEnerCon}. Upon fitting these terms
to the operators ${\cal O}_{\bf G} = \Delta \rho_{\bf G}^*$, we
may construct the function 
$v_m^{(2)}(\mbox{\boldmath $\alpha$};\mbox{\boldmath $\alpha$}^0)$.
This defines the fitted terms $T_{\bf G}$ in the predictor 
(Sec.\ \ref{Sec.IterProc}).

For a given ${\bf q}$ in BZ, we use the expressions
for $\epsilon^{(2)}$ and the derivatives 
$\partial \epsilon^{(2)} / \partial u_{{\bf q+G},{\bf q+G'}}$,
given in Appendix\ \ref{App.2bEnerCon},
corresponding to the two-body Jastrow factor $J_{\rm ih}({\bf q})$ 
defined above.
We construct the function $v^{(2)}_{{\bf q+G},{\bf q+G'}}( {\bf u_q} )$,
where 
${\bf u_q} = \{ u_{{\bf q+G},{\bf q+G'}} ; ~\mbox{for}~ {\bf G},{\bf G'} \}$,
and define the numerical predictor terms
$T_{{\bf q+G},{\bf q+G'}} = v^{(2)}_{{\bf q+G},{\bf q+G'}}$,
which contribute to the linear dependence
of the predictor on ${\bf u_q}$.

% ------------------------------------------------------------------------
\section{Results}
\label{Sec.Results}
% ------------------------------------------------------------------------

We now apply this optimization method to diamond and rhombohedral graphite. 
We shall compare the correlation factors determined in both these systems,
given the inhomogeneity and anisotropy  
of the electron charge density in graphite, relative to diamond.

The convergence criterion of our iterative optimization:
$V_m(\mbox{\boldmath $\alpha$}^n) = 0$,
requires the examination of possibly thousands of parameters,
and it is difficult to visualize the overall convergence of the method.
For illustrative purposes, we use the coefficients 
$V_m(\mbox{\boldmath $\alpha$}^n)$ 
to construct a single function $V^n$.
In the chosen Fourier basis, this amounts to reconstructing 
the real-space function $V^n$ from its Fourier coefficients. 
For one-body optimizations
\begin{eqnarray}
  V^n({\bf r}) = \sum_{\bf G} V_{\bf G}(\mbox{\boldmath $\chi$}^n)
                              e^{i{\bf G \cdot r}}~,
  \label{eq:1bodyV}
\end{eqnarray}
and for two-body optimizations
\begin{eqnarray}
  \label{eq:2bodyV}
  \lefteqn{V^n({\bf r},{\bf r'}) =} \hspace{0.5cm} & &
  \\ 
  & & \sum_{\bf q} \sum_{\bf G G'} 
                          e^{-i ({\bf q+G}) {\bf \cdot r}}
                          V_{{\bf q+G},{\bf q+G'}}({\bf u}^n)
                          e^{i ({\bf q+G'}) {\bf \cdot r'}}~.
  \nonumber
\end{eqnarray}

\begin{figure}
\begin{center}
  \leavevmode
  \begin{minipage}{4.00cm}
    \includegraphics[width=3.8cm]{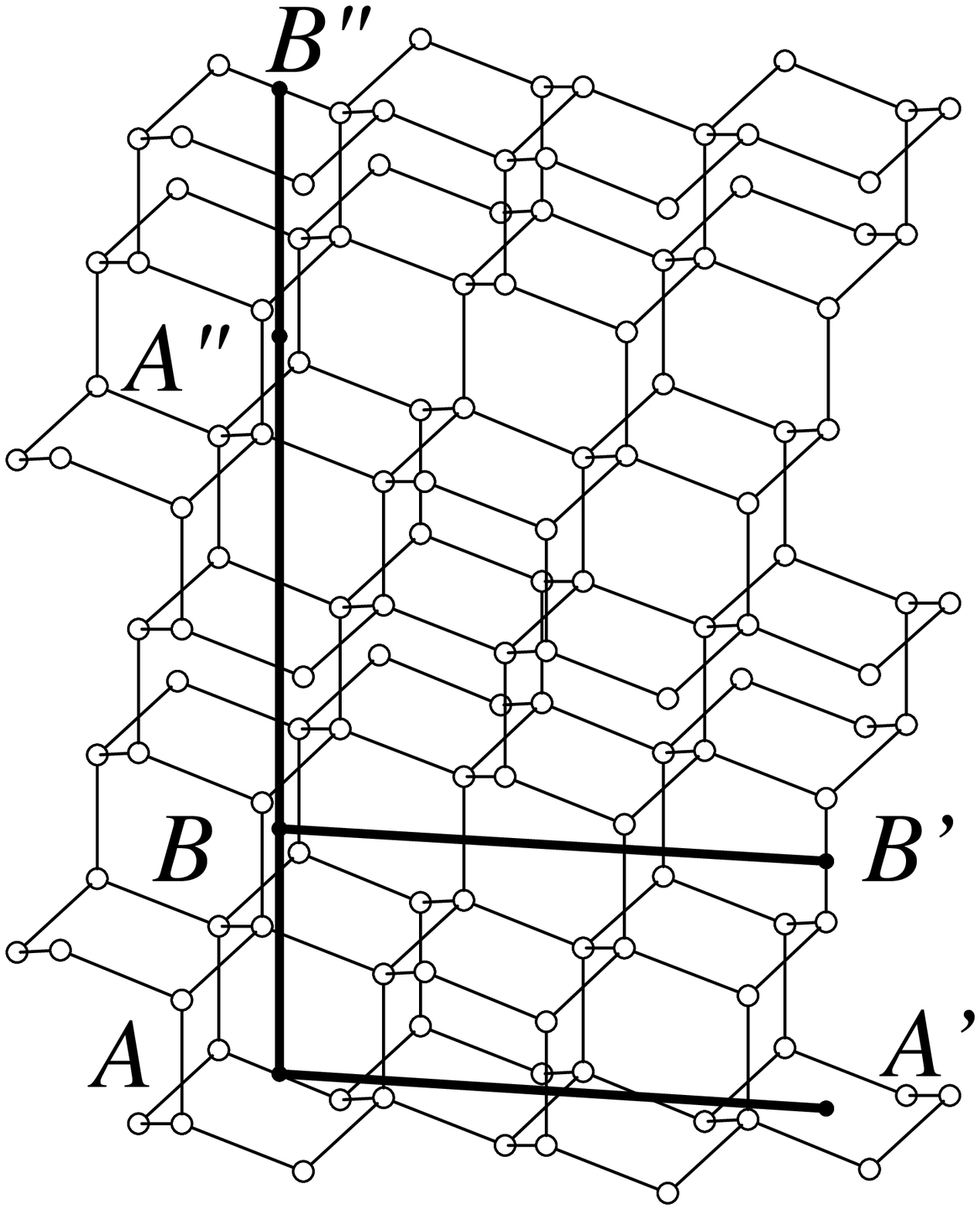}
  \end{minipage}
  \begin{minipage}{4.00cm}
    \includegraphics[width=4.2cm]{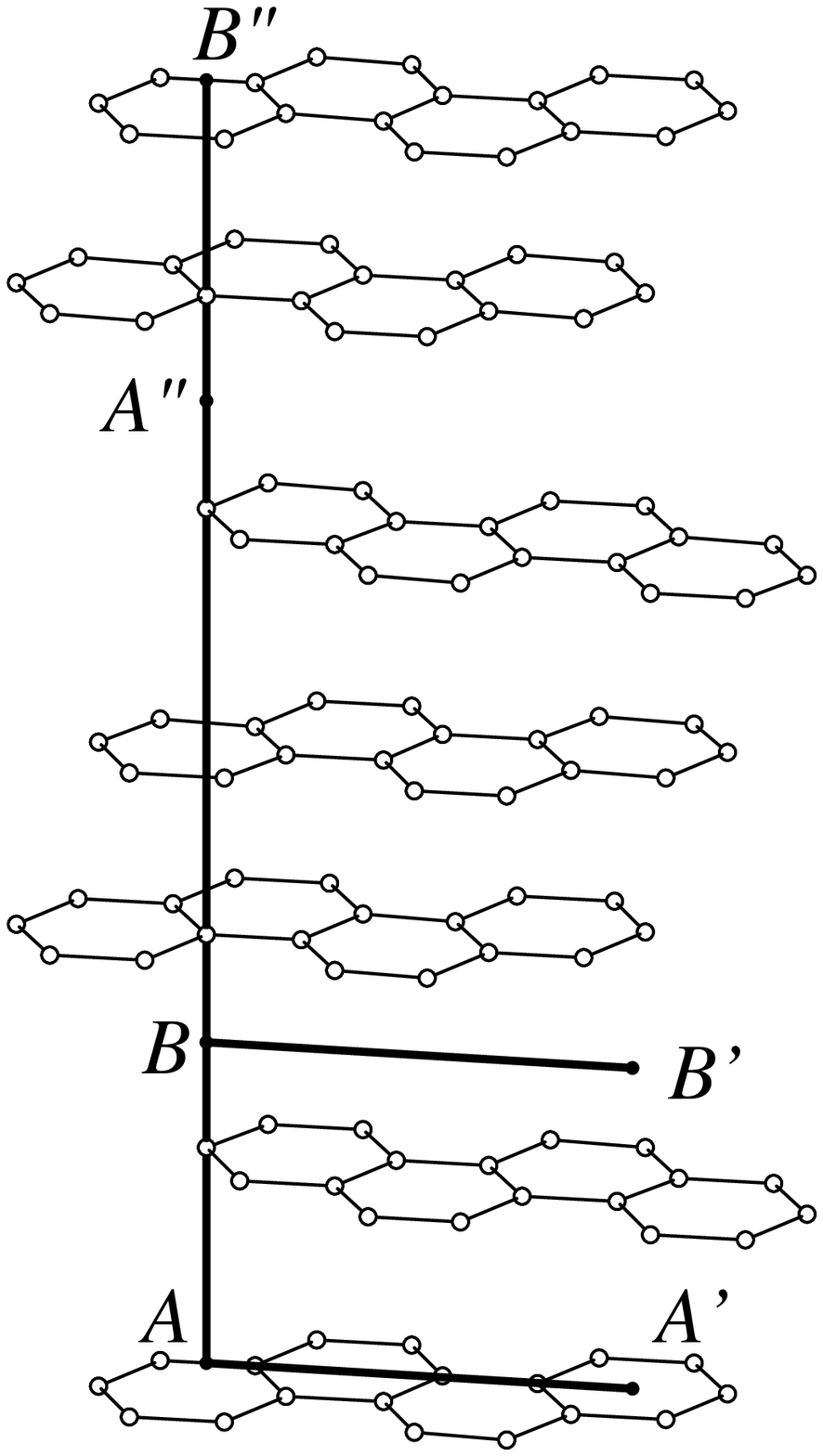}
  \end{minipage}
  \vspace{0.3cm}
  \begin{minipage}{4.00cm}
    \begin{center}
      (a) diamond
    \end{center}
  \end{minipage}
  \begin{minipage}{4.00cm}
    \begin{center}
      (b) graphite
    \end{center}
  \end{minipage}
  \caption{
    Crystal structure of (a) diamond and (b) rhombohedral graphite, 
    illustrating stacked layers of hexagonally arranged carbon atoms
    for graphite and buckled layers for diamond.
    In both structures, the point $A$ lies at the midpoint of a carbon-carbon
    bond, with the lines $AA'$ and $AA''$ extending within a layer and 
    perpendicular to the layers, respectively.
    The point $B$ lies midway between two layers (at a hexagonal interstitial
    point in diamond) and the lines $BB'$ and $BB''$ extend between the layers
    and perpendicular to the layers, respectively.
  }
  \label{fig:superlattice}
\end{center}
\end{figure}

Two-body functions, such as $V^n({\bf r},{\bf r'})$ 
and the Jastrow correlation function $u({\bf r},{\bf r'})$,
are functions of six variables and so, extracting useful information from
them is difficult.
For illustrative purposes, we indicate in Fig.\ \ref{fig:superlattice} 
two points, $A$ and $B$, in both the diamond 
and rhombohedral graphite structures,
corresponding to high and low electron charge density regions, respectively.
$A$ lies midway between two bonded carbon atoms and $B$ lies midway between
two layers of carbon atoms. 
We shall position the first electron at either $A$ or $B$.
The second electron shall be moved away from this position along one of
the following line segments
(indicated by heavy black lines in Fig.\ \ref{fig:superlattice}): 
  $AA'$ lying within a layer of carbon atoms;
  $AA''$ perpendicular to the layers;
  $BB'$ lying between two layers; and $BB''$ perpendicular to the layers.

By this means we may plot inhomogeneous two-body functions 
in terms of the relative separation of the electrons in the system.
In particular, we may draw some conclusions about electron correlation
in the system by examination of $u$.
We may determine the isotropy of $u$ by comparing
plots of $u$ with the first electron kept at the same point but the second
electron moved in perpendicular directions, 
e.g., by comparing plots designated by $AA'$ and $AA''$.
The homogeneity of $u$ may be seen by comparing plots of $u$ with the
second electron moving in parallel directions from different positions of
the first electron, e.g., by comparing $AA'$ and $BB'$.
Any differences between these plots of $u$ are attributable to the
inhomogeneity and anisotropy of the electron correlation factors in the
systems studied. 

We use periodic boundary conditions (PBC) to approximate 
the infinite crystal.\ \cite{FahyWang} 
The simulation cells consist of an $N_1 \times N_2 \times N_3$ unit cell
arrangement. The unit cells in each system are defined by 
the Bravais lattice basis vectors. 
For diamond, we use the basis: 
$\{ (a/2,a/2,0); \ (0,a/2,a/2); \ (a/2,0,a/2) \}$,
where $a=6.72$ a.u., corresponding to a carbon bond-length of
$2.91$ a.u.
For rhombohedral graphite we use the basis: 
$\{ (0,a,c); \ (-\sqrt{3}a/2,-a/2,c); \ (\sqrt{3}a/2,-a/2,c) \}$, 
where $a=2.68$ a.u. is the bond-length within the layers,
and $c=6.33$ a.u. is the layer separation.

We construct the Slater determinant $D$ for both systems using
DFT calculations in the local density approximation (LDA).\ \cite{KohnSham}
The LDA orbitals were expanded using a linear combination 
of atomic orbitals 
comprising gaussians centred on each 
of the two carbon atoms in 
the unit cell.\ \cite{FahyWang,Malatesta,ChanVanderbilt}

% ------------------------------------------------------------------------
\subsection{Removal of cusp}
\label{Sec.RemCusp}
% ------------------------------------------------------------------------

\begin{figure}
\vspace{0.2cm}
\begin{center}
  \leavevmode
  \includegraphics[angle=-90,width=8.0cm]{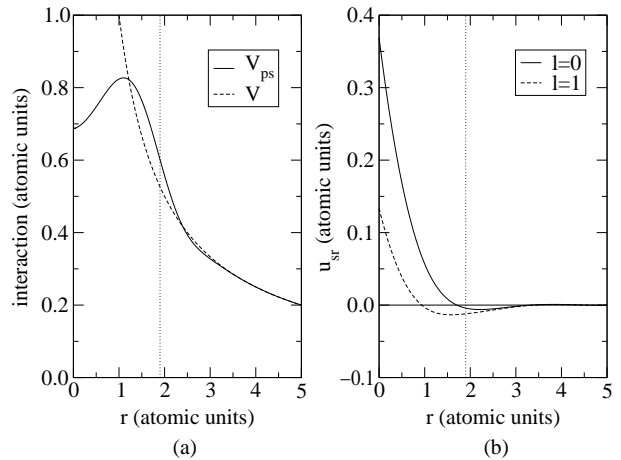}
  \caption{
    (a) The pseudo-interaction $V_{\rm ps}$ (solid line)
        and the Coulomb interaction $V=1/r$ (dashed line)
        versus electron separation $r$.
    (b) The short-range Jastrow function $u_{\rm sr}$,
        generated from $V_{\rm ps}$,
        versus electron separation $r$, for angular momenta
        $l=0$ (solid line) and $l=1$ (dashed line).
    $V_{\rm ps}$ is generated using $r_c = 1.9$ a.u. 
    (indicated by vertical dotted line)
    and $\epsilon = 0.2$ hartree (see text).
  }
  \label{fig:pseudousr}
\end{center}
\end{figure}

In Sec.\ \ref{Sec.ElecCusp}, we discussed the advantages of removing
the short-range cusp from the function $u$ in order to
improve its representability as a linear
combination of smooth functions.
Figure\ \ref{fig:pseudousr}(a) compares the pseudointeraction $V_{\rm ps}$, 
generated using a cut-off radius of $r_c = 1.9$ a.u. 
and energy eigenvalue $\epsilon = 0.2$ hartree (as discussed in
Ref.~\onlinecite{Cusppaper}), 
with the Coulomb interaction $V = e^2/r$. (In atomic units $e^2=1$).
The pseudointeraction is used to generate a 
short-range Jastrow funtion $u_{\rm sr}$, which is shown in
Fig.\ \ref{fig:pseudousr}(b)  for the relative angular momenta
$l=1$ and $l=0$ corresponding to parallel spin and anti-parallel spin 
correlation, respectively. 
The short range Jastrow factor used in all subsequent 
calculations is that generated with these particular values of $r_c$
and $\epsilon$. Subsequent figures in this paper, 
which involve $u_{\rm sr}$, represent
anti-parallel spin correlation only.

\begin{figure}
\begin{center}
  \leavevmode
  \includegraphics[angle=-90,width=8.0cm]{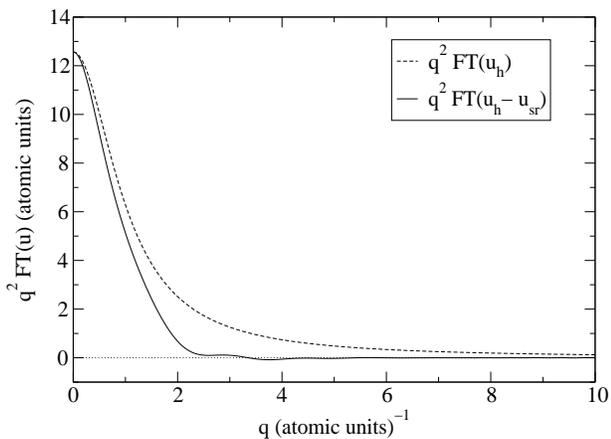}
  \caption{
    Fourier transform times $q^2$ of 
    (a) the homogeneous Jastrow factor $u_{\rm h}$ (dashed line) and
    (b) the cuspless difference $u_{\rm h} - u_{\rm sr}$ (solid line)
    versus wave vector $q$. The homogeneous Jastrow factor parameter
    $A = 1$ a.u. (Eq.\ \ref{eq:Yukawa}).
  }
  \label{fig:cuspremove}
\end{center}
\end{figure}

The cut-off required for a convergent Fourier expansion 
of a smooth cuspless function should be much less than that required for  
a function with a short-range cusp. 
Therefore, using a cuspless form greatly reduces
the number of terms required to represent the inhomogeneous form of 
the Jastrow factor accurately in Fourier space. 
We illustrate this point using a simple example. 
In Fig.\ \ref{fig:cuspremove} we plot $q^2$ times the Fourier transform, 
for wave vector $q$, of the Yukawa-style homogeneous correlation factor,
\begin{eqnarray}
  u_{\rm h} = \frac{A}{r} ( 1 - e^{-r/F} )~,
  \label{eq:Yukawa}
\end{eqnarray}
which has been used by many authors to approximate electron correlation in
a variety of systems.\ \cite{FahyWang,1bodyEFO,Bevan,Gaudoin,Malatesta,Ceperley}
We set $A=1$ a.u., 
($F$ is determined from $A$ to satisfy the cusp conditions\ \cite{KatoCusp} 
and depends on the relative spin of the electrons).
Also shown in Fig.\ \ref{fig:cuspremove} is $q^2$ times the 
Fourier transform of the cuspless difference $u_{\rm h} - u_{\rm sr}$.
We assume that, for large electron separations, electron
correlation is approximately spin-independent. Therefore, for each $q$,
we plot the mean of the parallel spin and anti-parallel spin values 
of the functions.

In practice, we use the first zero of the Fourier transform of
$V_{\rm ps}$ as the Fourier space cut-off $k_c$ used in 
the definition of $J_{\rm ih}$ in Eq.\ \ref{eq:inhomJast}.
For $r_c=1.9$ a.u. we use the cut-off $k_c=2.185 ~(\mbox{a.u.})^{-1}$,
beyond which the Fourier transform of the
cuspless $u$ function is approximately zero (Fig.\ \ref{fig:cuspremove}).
Combining both the short-range and inhomogeneous forms
of the Jastrow factor using this scheme produces a form that is 
approximately independent of the cut-off, since decreasing $r_c$ increases
the reciprocal space cut-off $k_c$.

% ------------------------------------------------------------------------
\subsection{Homogeneous Jastrow Factor}
\label{Sec.HomoJ}
% ------------------------------------------------------------------------

\begin{figure}[!]
\begin{center}
  \leavevmode
  \vspace{0.2cm}
  \includegraphics[angle=-90,width=8.0cm]{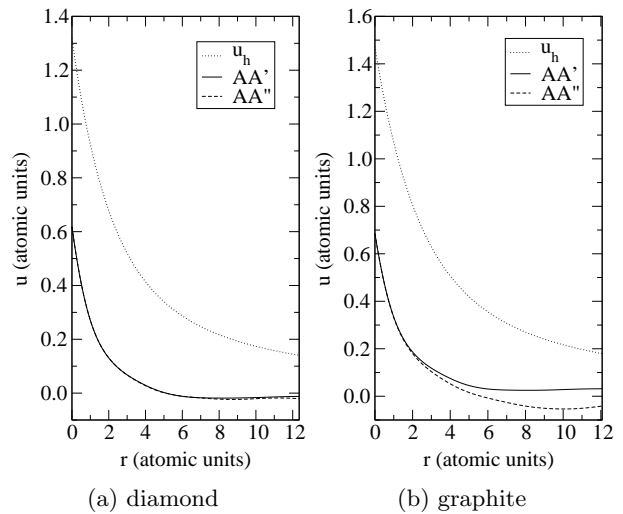}
  \begin{minipage}{4.0cm}
    \center{(a) diamond}
  \end{minipage}
  \begin{minipage}{4.0cm}
    \center{(b) graphite}
  \end{minipage}
  \caption{
    Jastrow correlation factors versus electron separation $r$ for
    (a) diamond and (b) rhombohedral graphite. 
    (i) The function $u_{\rm h}$ (dotted line) as defined in 
    Eq.\ \ref{eq:Yukawa} with optimized parameter $A = 1.739$ for diamond 
    and $2.170$ for graphite.
    (ii) The reconstructed function 
    $u$ of Eq.\ \ref{eq:Yukawarecon}, from
    a $3 \times 3 \times 3$ simulation region in both systems, for
    electron separations along the line segments
    $AA'$ (solid line) and $AA''$ (dashed line) 
    from Fig.\ \ref{fig:superlattice}.
  }
  \label{fig:uhom}
\end{center}
\end{figure}

\begin{figure*}[t]
\begin{center}
  \leavevmode
  \includegraphics[angle=-90,width=12.0cm]{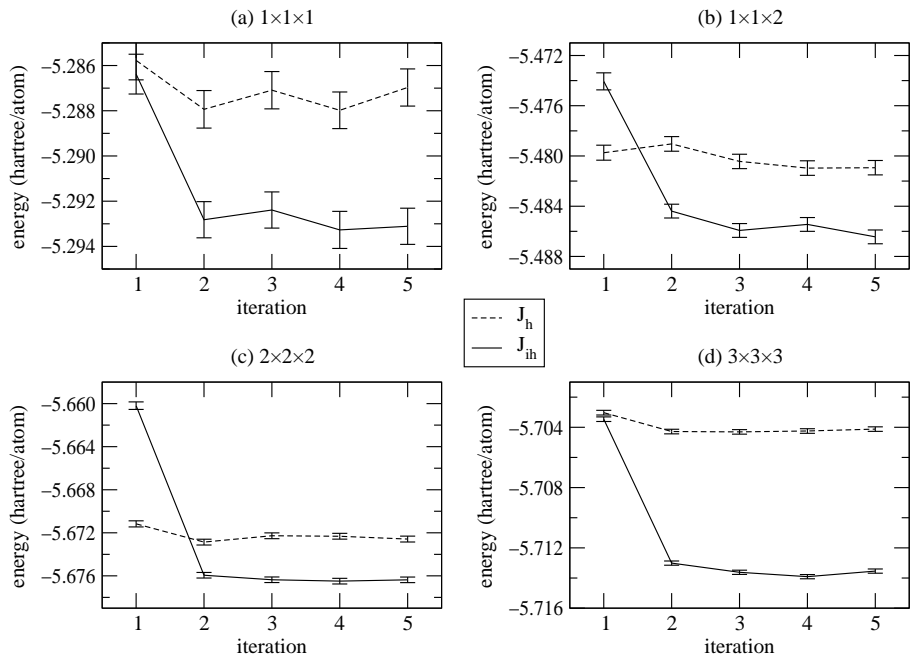}
  \vspace{0.2cm}
  \caption{
    Energy of diamond, in hartree/atom, 
    for each iteration of the optimization process
    and various simulation cell sizes using the inhomogeneous Jastrow factor
    $J_{\rm ih}$ (solid line): 
    (a) $1 \times 1 \times 1$ unit cell, with $N=8$ electrons and 
        $N_{\alpha}=140$ variational wave function parameters;
    (b) $1 \times 1 \times 2$, $N=16$, $N_{\alpha}=196$; 
    (c) $2 \times 2 \times 2$, $N=64$, $N_{\alpha}=532$; 
    (d) $3 \times 3 \times 3$, $N=216$, $N_{\alpha}=1235$.
    Also shown (dashed line) are energies during optimization 
    of the one-body Jastrow factor
    $J_{\rm 1b}$ (with 84 variational parameters) in combination with the 
    homogeneous Jastrow factor $J_{\rm h}$. 
    Averages for each iteration 
    are calculated using $10^5$ Monte Carlo samples.
  }
  \label{fig:diamondener}
\end{center}
\end{figure*}

For comparison with the inhomogeneous $u$ functions determined in
the following sections we use the Yukawa-style homogeneous 
function $u_{\rm h}$ of Eq.\ \ref{eq:Yukawa} and construct a homogeneous trial
wave function of the form $\Psi = J_{\rm sr} J_{\rm h} J_{\rm 1b} D$.
As with the inhomogeneous trial wave function, we represent short-range
correlation (i.e., the cusp) using $J_{\rm sr}$ and represent the remaining
correlation using a homogeneous Jastrow factor
\begin{eqnarray*}
  J_{\rm h} = \exp \left[- \sum_{q<k_c} \tilde{u}({\bf q}) P_{\bf q q} \right]~,
\end{eqnarray*}
where we define
\begin{eqnarray*}
  \tilde{u}({\bf q}) \equiv \int_{\Omega} [u_{\rm h}(r) - u_{\rm sr}(r)]
                            e^{-i{\bf q \cdot r}} d{\bf r}~.
\end{eqnarray*}

For the uniform electron gas, 
Bohm and Pines\ \cite{BohmPines} predicted that
the true function $u$ should decay as $1/\omega_p r$,
at large separation $r$, where $\omega_p$ is the
plasma frequency. Rather than use this limiting value $A= 1/\omega_p$ in
Eq.\ \ref{eq:Yukawa}, 
it is common to treat $A$ as a free
parameter such that the energy is minimized.
Using variational calculations we can
determine the optimal value of $A$.\ \cite{Malatesta} 
We optimize the one-body Jastrow factor 
$J_{\rm 1b}$ in Eq.\ \ref{eq:1bJast}, using our iterative method.
The Jastrow factor $J = J_{\rm sr} J_{\rm h} J_{\rm 1b}$, 
combined with the Slater determinant $D$, 
is our best approximation of the true many-body
eigenstate using a homogeneous two-body Jastrow factor 
and is comparable with similar wave functions optimized
using variance minimization.\ \cite{FahyWang,Gaudoin,Malatesta} 

In Fig.\ \ref{fig:uhom} we plot $u_{\rm h}$ and compare it with 
the reconstructed function
\begin{eqnarray}
  u({\bf r}) = u_{\rm sr}(r) + \sum_{q<k_c} \tilde{u}({\bf q}) 
               e^{i{\bf q \cdot r}}~, 
  \label{eq:Yukawarecon}
\end{eqnarray}
for $3 \times 3 \times 3$ unit cell
simulations of diamond and rhombohedral graphite. 
Periodic boundary conditions and the anisotropy of the unit cell 
make this reconstructed form
appear quite different to the original isotropic function
$u_{\rm h}$. 
 
In particular, for rhombohedral graphite 
the unit cell used is quite anisotropic, leading to marked differences in the
reconstructed function along the perpendicular line segments 
$AA'$ and $AA''$.
(The homogeneity of $u_{\rm h}$ is preserved
and so we only plot the function for point $A$, since all other 
points are equivalent.)
Note that the Jastrow $u$ function is defined 
up to an arbitrary constant, much like
a potential, since this constant affects only 
the wave function normalization and
contributes nothing to the description of correlation. 
Therefore, it is of no consequence that the functional
form of $u_{\rm h}$ in Eq.\ \ref{eq:Yukawa} appears shifted above 
the reconstructed forms compatible with PBC. 
This is due to the removal of the constant Fourier coefficient of the
correlation function $u({\bf G=0})$ from the expansion of the
reconstructed function in Eq.\ \ref{eq:Yukawarecon}.
We note that the cusp conditions\ \cite{KatoCusp}
are maintained by all forms.

We optimize the one-body Jastrow $J_{\rm 1b}$
using the method described in
Sec.\ \ref{Sec.EnerMin}. 
For the diamond simulations we used a Fourier space cut-off
$G_c = 5.0~\mbox{(a.u.)}^{-1}$, giving 84 variational one-body parameters.
This is more than enough for an accurate representation of one-body
terms in the wave function (see Sec.\ \ref{Sec.EnergyCorrections}).
For graphite simulations we used
$G_c = 3.1~\mbox{(a.u.)}^{-1}$, giving 30 variational 
one-body parameters.

\begin{figure*}[t]
\begin{center}
  \leavevmode
  \includegraphics[angle=-90,width=12.0cm]{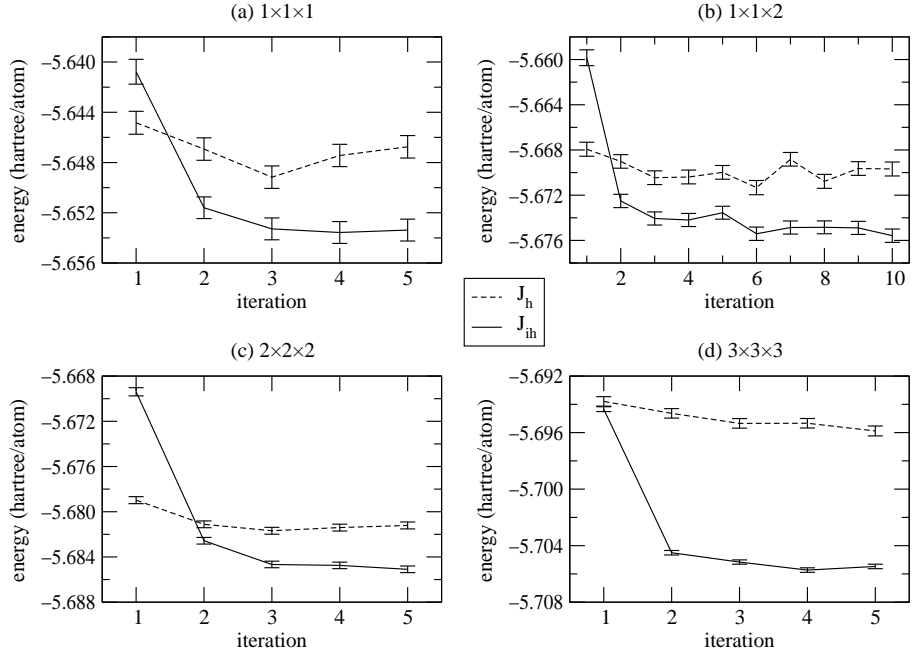}
  \vspace{0.2cm}
  \caption{
    Energy of rhombohedral graphite, in hartree/atom, 
    for each iteration of the optimization process 
    and various simulation cell sizes using the inhomogeneous Jastrow factor
    $J_{\rm ih}$ (solid line): 
    (a) $1 \times 1 \times 1$ unit cell, with $N=8$ electrons and 
        $N_{\alpha}=162$ independent parameters for $J_{\rm ih}$;
    (b) $1 \times 1 \times 2$, $N=16$, $N_{\alpha}=272$; 
    (c) $2 \times 2 \times 2$, $N=64$, $N_{\alpha}=1020$; 
    (d) $3 \times 3 \times 3$, $N=216$, $N_{\alpha}=3136$.
    Also shown (dashed line) are energies during optimization 
    of the one-body Jastrow factor
    $J_{\rm 1b}$ (with 30 variational parameters) in combination with the 
    homogeneous Jastrow factor $J_{\rm h}$. 
    Averages for each iteration 
    are calculated using $10^5$ Monte Carlo samples.
    (The optimization of $J_{\rm 1b}$ for the 
    $3 \times 3 \times 3$ simulation used only $2.5 \times 10^4$ samples
    per iteration.)
  }
  \label{fig:graphiteener}
\end{center}
\end{figure*}

The values of the electronic energy per atom for various optimizations
of this homogeneous trial wave function are shown in 
Figs.\ \ref{fig:diamondener} and\ \ref{fig:graphiteener}.
Each iteration involved averaging over $10^5$ Monte Carlo samples.
However, this amount of averaging was more than enough for an accurate
implementation of the method.
The optimization of the one-body terms in the $3 \times 3 \times 3$
simulation of graphite (Fig.\ \ref{fig:graphiteener}) 
involved only $2.5 \times 10^4$ samples per iteration and is well converged.
On average, the gain in energy following one-body optimization is
approximately $1.0$ mhartree/atom for diamond and
$2.5$ mhartree/atom for graphite. 
We note that the necessity for a one-body correction is a consequence of
the inhomogeneity of the electronic charge density in the 
system,\ \cite{Malatesta,Gaudoin} and it is not surprising that the gain
in energy is larger for the more inhomogeneous system, graphite,
than for diamond.

The Slater determinant used in the diamond calculations of 
Fig.\ \ref{fig:diamondener} is composed of single-particle orbitals
generated from an LDA calculation. The LDA orbitals are linear combinations
of gaussian basis functions of $s$, $p$ and $d$
symmetry, using three decays of 0.24, 0.797, and 2.65. 
The exchange-correlation functional used
was of the Ceperley-Alder\ \cite{CepAldDMC} form. The 
cut-off of the Fourier space expansion of the 
charge density\ \cite{ChanVanderbilt} in the LDA calculation was 64 Rydberg. 

The graphite calculations shown in Fig.\ \ref{fig:graphiteener} use a
Slater determinant generated using a 
gaussian basis-set with $s$ and $p$ symmetry only.
Four decays were used: 0.19, 0.474, 1.183, and 2.95.
The orbitals were generated from LDA calculations incorporating
the Hedin-Lundqvist\ \cite{HedLundXC} exchange-correlation functional,
and a cut-off of 36 Rydberg for the Fourier space expansion of the
charge density. (See Sec.\ \ref{Sec.EnergyCorrections} for a discussion
of basis-set convergence of the total energy in graphite.)

\begin{figure}
\begin{center}
  \leavevmode
  \includegraphics[angle=-90,width=8cm]{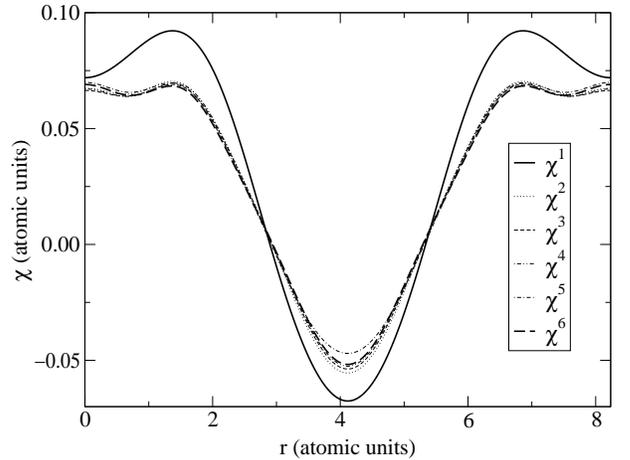}
  \caption{
    The diamond one-body function $\chi$ versus position $r$ along the line
    segment $AA'$ [Fig.\ \ref{fig:superlattice}(a)].
    $\chi^n$ indicates the one-body function used at iteration $n$
    of the optimization of $J_{\rm 1b}$ in the presence of $J_{\rm h}$ in
    Fig.\ \ref{fig:diamondener}(d).
  }
  \label{fig:1boptchi}
\end{center}
\end{figure}

Figure\ \ref{fig:1boptchi} illustrates the convergence during optimization 
of the one-body function $\chi({\bf r})$ in Eq.\ \ref{eq:JastrowPsi}
(where $\chi$ is the accumulation of all one-body terms 
from each of the Jastrow factors)
for the $3\times 3 \times 3$ diamond calculation. 
The optimized function is statistically well-determined, and 
the change from the initial one-body function, $\chi^1=\chi^0$
of Eq.\ \ref{eq:natural1body} 
(as defined in Refs.~\onlinecite{Gaudoin} and~\onlinecite{Malatesta}), 
is well-defined.  This alteration of the one-body function
may be compared to similar calculations performed using variance
minimization.\ \cite{Malatesta,Gaudoin}

The number of parameters for optimization could be greatly reduced
through exploitation of the crystal point-group symmetry 
of the structures involved.
However, it is worth noting that the optimization process preserves
the natural symmetry of the system (within statistical error)
without such measures,
illustrating that for non-symmetric systems with large numbers of parameters,
this optimization process should be quite robust.

\begin{figure}
\begin{center}
  \leavevmode
  \includegraphics[angle=-90,width=8cm]{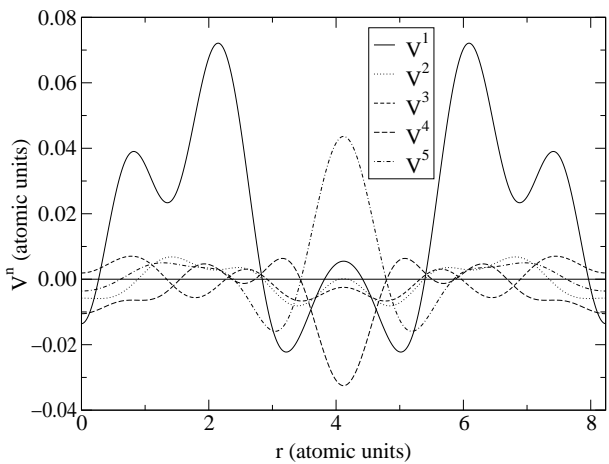}
  \caption{
    The reconstructed one-body function $V^n$, defined in 
    Eq.\ \ref{eq:1bodyV}, versus position $r$ as for 
    Fig.\ \ref{fig:1boptchi}. 
  }
  \label{fig:1boptEL}
  \vspace{0.5cm}
\end{center}
\end{figure}

The one-body function $V^n({\bf r})$, reconstructed from the coefficients
associated with the local energy $V_{\bf G}$ at iteration $n$ according to 
Eq.\ \ref{eq:1bodyV}, is shown in
Fig.\ \ref{fig:1boptEL}. Clearly, this function decreases in magnitude,
indicating a decrease in the magnitudes of the Euler-Lagrange derivatives.
Beyond the first iteration, $V^n({\bf r})$ is of the same order of
magnitude as its associated standard error, and so is statistically 
insignificant.
Therefore, the method has essentially converged after only one iteration. 
The noisiest regions of $V^n({\bf r})$ correspond to 
regions of low density where the Monte Carlo sampling is less frequent.

% ------------------------------------------------------------------------
\subsection{Inhomogeneous RPA Jastrow Factor}
\label{Sec.inhomRPAJ}
% ------------------------------------------------------------------------

\begin{figure}
\begin{center}
  \leavevmode
  \includegraphics[angle=-90,width=8cm]{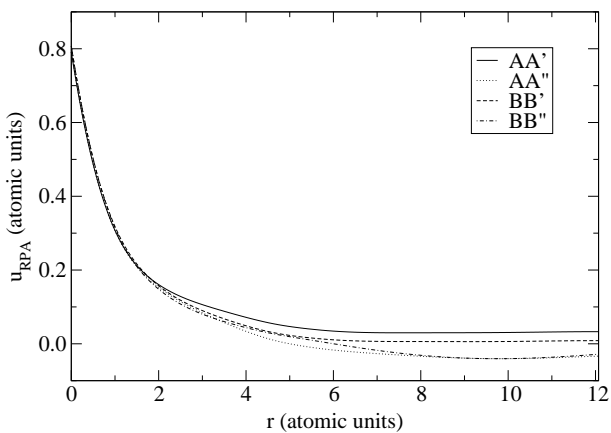}
  \caption{
    The Jastrow factor $u_{\rm RPA}$ versus electron separation $r$
    along the line segments indicated in Fig.\ \ref{fig:superlattice}(b),
    for a $3 \times 3 \times 3$ simulation of rhombohedral graphite.
  }
  \label{fig:uRPA}
\end{center}
\end{figure}

The analytic guess for the two-body predictor functions 
$S_{\bf q q'}({\bf u})$,
outlined in Sec.\ \ref{Sec.PredictorAnalApprox.J2b}, leads
to an inhomogeneous generalization of the RPA equations.
The solution to these equations is the function $u_1 = u_{\rm RPA}$,
shown in Fig.\ \ref{fig:uRPA} for the $3 \times 3 \times 3$ 
graphite simulation. 
We notice some inhomogeneity in $u_{\rm RPA}$ at 
intermediate- and long-range electron separations.
A more homogeneous and isotropic $u_{\rm RPA}$
was found for diamond, as we would expect 
since diamond possesses a more uniform electron density than graphite.

In Figs.\ \ref{fig:diamondener} and\ \ref{fig:graphiteener}, 
the first point on all solid curves indicates the total energy per atom
in each simulation calculated using $u_{\rm RPA}$. 
In comparison with the energy calculated using 
the homogeneous Jastrow function $u_{\rm h}$,
we see that the inhomogeneous RPA trial wave function 
is at best comparable in accuracy with the optimized 
trial wave function with homogeneous two-body Jastrow factor, 
and often less accurate.
These results are different from those of
Gaudoin {\it et al.}\ \cite{Gaudoin}  
for model systems: they find that their inhomogeneous generalization of
the RPA produces wave functions that yield lower energies
than the homogeneous form.

% ------------------------------------------------------------------------
\subsection{Inhomogeneous Optimal Jastrow Factor}
\label{Sec.inhomJopt}
% ------------------------------------------------------------------------

We simultaneously optimized the parameters in both the one-body Jastrow factor
$J_{\rm 1b}$ and the fully inhomogeneous form of the two-body Jastrow factor
$J_{\rm ih}$, using our iterative method. The convergence of the total energy
per atom for various simulations of diamond and rhombohedral graphite are 
shown in Figs.\ \ref{fig:diamondener} and\ \ref{fig:graphiteener}. 
(The Slater determinants used in combination with the homogeneous
Jastrow factor $J_{\rm h}$ in Sec.\ \ref{Sec.HomoJ} are also used here.)
Convergence of the total energy is achieved in approximately three iterations
in all cases and is stable. 
This is remarkable given that the system sizes range from
8 electrons and 140 independent parameters to 216 electrons and 3136 
independent parameters. Also, we used the same amount of Monte Carlo sampling,
viz., $10^5$ samples per iteration, 
to determine the required expectation values for all simulations.
In all cases the fully inhomogeneous form of $u$
allows us to determine more accurate trial wave functions with 
substantially lower energies
than the homogeneous trial functions. In general, the gain in energy through
using an inhomogeneous rather than a homogeneous wave function is 
of the order of $5$ mhartree/atom for both diamond and rhombohedral graphite.

\begin{figure}
\begin{center}
  \leavevmode
  \includegraphics[angle=-90,width=8cm]{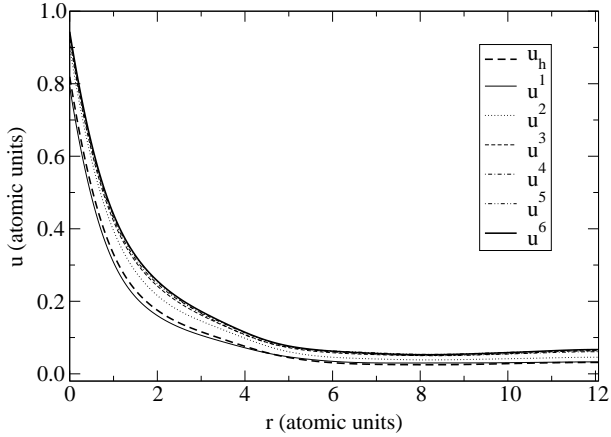}
  \caption{
    Graphite Jastrow correlation functions $u^n$ with respect to 
    electron separation $r$ on the line segment $AA'$ 
    (Fig.\ \ref{fig:superlattice}) for each iteration $n$ during optimization of
    $J_{\rm ih}$ for a $3 \times 3 \times 3$ simulation region
    [Fig.\ \ref{fig:graphiteener}(d)].
    Also shown is the reconstruction of the homogeneous function $u_h$ 
    (heavy dotted line) defined in Eq.\ \ref{eq:Yukawarecon} with
    optimized parameter $A = 2.170$.
  }
  \vspace{0.2cm}
  \label{fig:graphuconv}
\end{center}
\end{figure}

\begin{figure}
\begin{center}
  \leavevmode
  \includegraphics[angle=-90,width=8cm]{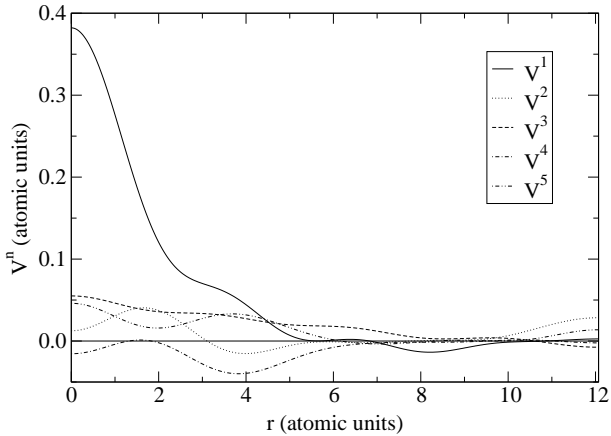}
  \caption{
    The reconstructed two-body function $V^n$, as defined in 
    Eq.\ \ref{eq:2bodyV}, versus electron separation $r$ on
    the line segment $AA'$, for each iteration $n$ during the optimization
    outlined in Figs.\ \ref{fig:graphiteener}(d) and\ \ref{fig:graphuconv}.
  }
  \label{fig:graphvconv}
\end{center}
\end{figure}

Figure\ \ref{fig:graphuconv} illustrates the rapid convergence of the
two-body wave function parameters ${\bf u}$ to their optimal values
during the largest graphite optimization ($3 \times 3 \times 3$) and is
typical of all the optimizations performed in both diamond and graphite.
Beyond the third iteration, no clear distinction exists between subsequent 
sets of parameters. 
The optimal Jastrow function $u$ is significantly different from 
both the RPA function $u^1$ and the homogeneous form $u_h$
of Eq.\ \ref{eq:Yukawa}. The proof that this optimization succeeds in
minimizing the energy expectation value may be seen in 
Fig.\ \ref{fig:graphvconv} (which comes from the same graphite calculation
as Fig.\ \ref{fig:graphuconv}). Here, we plot the iterative decay of the
two-body function $V^n({\bf r},{\bf r'})$ reconstructed from the 
total energy coefficients determined at each iteration,
according to Eq.\ \ref{eq:2bodyV}. This
clearly indicates the reduction to zero (within statistical noise)
of the derivatives in the Euler-Lagrange equations, thus solving the
energy minimization problem.

\begin{figure}
\begin{center}
  \leavevmode
  \includegraphics[angle=-90,width=8cm]{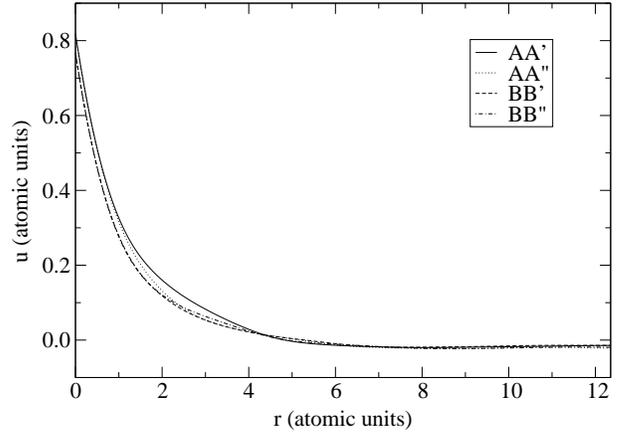}
  \caption{
    The optimized diamond Jastrow correlation factor $u$ as a
    function of electron separation $r$ along the line segments indicated,
    for a $3 \times 3 \times 3$ simulation region. 
  }
  \vspace{0.2cm}
  \label{fig:diamuopt}
\end{center}
\end{figure}

\begin{figure}
\begin{center}
  \leavevmode
  \includegraphics[angle=-90,width=8cm]{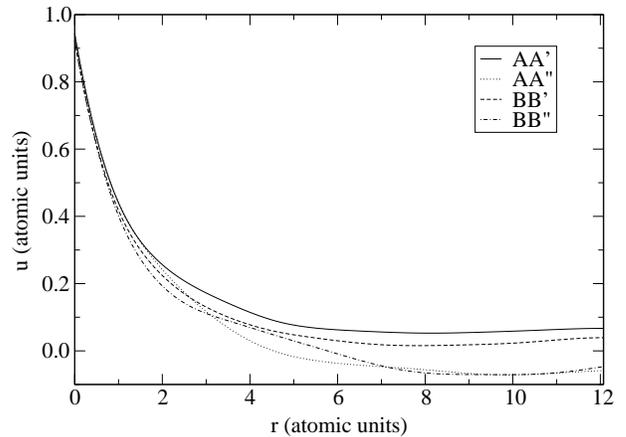}
  \caption{
    The optimized rhombohedral graphite Jastrow correlation factor $u$ as a
    function of electron separation $r$ along the line segments indicated,
    for a $3 \times 3 \times 3$ simulation region.
  }
  \label{fig:graphuopt}
\end{center}
\end{figure}

For the largest simulation cells studied 
($3 \times 3 \times 3$ unit cell arrangement containing 216 electrons), 
we compare the optimal Jastrow correlation functions $u$
of diamond and rhombohedral graphite. 
Figures\ \ref{fig:diamuopt} and\ \ref{fig:graphuopt} show the function $u$
plotted with respect to electron separation on various 
line segments in the corresponding crystal structures,
as already explained.

% ------------------------------------------------------------------------
\section{Discussion}
\label{Sec.Discussion}
% ------------------------------------------------------------------------

% ------------------------------------------------------------------------
\subsection{Energies}
\label{Sec.EnergyCorrections}
% ------------------------------------------------------------------------

For a direct comparison of the calculated energies of diamond and
rhombohedral graphite, we should (a) make some corrections based on
the trial wave functions used and (b) include finite size
and zero-point phonon energy corrections
for the expected energy of the real infinite solid.
The corrected energies for diamond and graphite are listed in
Table\ \ref{tab:enercorr}.

\begin{table}

\caption{
  Energies and energy corrections of 
  diamond and rhombohedral graphite (in hartree/atom). 
  (a) Total energy, determined by VMC and our optimization method,
  for a $3 \times 3 \times 3$ unit cell simulation region, using
  wave functions of similar variational freedom (see text).
  (b) Finite size correction equal to the energy difference between
  an LDA calculation for the $3 \times 3 \times 3$ simulation region 
  and an LDA calculation using a fully-converged k-point set (see text).
  (c) Correction for the zero-point phonon energy.\ \cite{zpe}
  (d) Total energy including the corrections.
  The numbers in parentheses indicate the statistical error in the last 
  digits of the corresponding energy.
  \label{tab:enercorr}
}
\begin{ruledtabular}
\begin{tabular}{lr@{.\hspace{-0.70cm}}lr@{.\hspace{-0.70cm}}l}
 & \multicolumn{2}{c}{diamond} & \multicolumn{2}{c}{graphite} \\
\hline
$3 \times 3 \times 3$  & $-5$&712\ 95(14) & $-5$&712\ 91(14) \\ 
finite size correction & $-0$&008\ 99     & $-0$&006\ 56     \\
zero-point energy      & $ 0$&006\ 65     & $ 0$&006\ 10     \\
total                  & $-5$&715\ 29(14) & $-5$&713\ 41(14) \\
\end{tabular}
\end{ruledtabular}
\end{table}

The Jastrow factors used in both solids are comparable in their 
variational freedom, the only significant
difference being the size of the cut-off used for $\chi_{\rm 1b}$.
However, a VMC calculation for diamond, using the same cut-off as
in graphite ($G_c = 3.1$ (a.u.)$^{-1}$, 
corresponding to 25 variational parameters), 
resulted in an increase in energy 
of only $1.0 \pm 0.3$ mhartree/atom for diamond. 
Therefore, the high Fourier coefficients of
$\chi_{\rm 1b}$ contribute little to the total energy of the system.
The diamond VMC energy for the $3 \times 3 \times 3$ simulation
quoted in Table\ \ref{tab:enercorr} was determined using 
$G_c = 3.1$ (a.u.)$^{-1}$ as the cut-off for one-body terms.

In graphite, the exclusion of $d$ symmetry from the basis set used to construct
the LDA orbitals in the Slater determinant
is energetically more important.
In addition to the calculations described in Secs.\ \ref{Sec.HomoJ} 
and\ \ref{Sec.inhomJopt}, we also performed calculations for
graphite using a gaussian basis set with $s$, $p$ and $d$
symmetry, and three gaussian decays: 0.22, 0.766, 2.67.
For the $3 \times 3 \times 3$ simulation, including $d$ symmetry reduces the
total VMC energy by $7.2 \pm 0.3$ mhartree/atom and also reduces 
the variance in the total energy by 16\%. 
Use of the Ceperley-Alder exchange correlation 
functional to generate the single-particle orbitals in graphite,
rather than the Hedin-Lundqvist form, made no
difference (within statistical error) to the VMC energies, and
neither did an increase in the cut-off for the Fourier space expansion of
the LDA charge density from 36 to 64 Rydberg.
The graphite VMC energy for the $3 \times 3 \times 3$ simulation
listed in Table\ \ref{tab:enercorr} was calculated using a 
trial wave function very similar to that used for the calculation of
the corresponding diamond VMC energy. The cut-off for the one-body 
Jastrow factor was $G_c = 3.1$ (a.u.)$^{-1}$ and the Slater determinant
comprised LDA orbitals obtained using: 
(i) a basis set with $d$ symmetry (as outlined above);
(ii) the Ceperley-Alder exchange correlation functional;
and (iii) a 64 Rydberg cut-off for the LDA charge density expansion.

We generate finite size corrections for the $3 \times 3 \times 3$ unit cell
simulations of diamond and graphite, by calculating the difference in energy
between an LDA calculation which uses k-points compatible with 
periodic boundary conditions of a $3 \times 3 \times 3$ simulation region 
and an LDA calculation using a fully converged
k-point set.\ \cite{FahyWang} 
Comparing the change in energy between a $2 \times 2 \times 2$ calculation
and a $3 \times 3 \times 3$ calculation in both diamond and graphite,
using LDA and VMC, we see that the change in VMC energy is about 80\%
of the LDA energy change in diamond, and 70\% in rhombohedral graphite.
Perhaps more accurate estimates of the energy of the infinite solid
may be obtained by implementation of a model periodic Coulomb interaction
developed recently. Tests of this approach have dramatically reduced
finite-size effects in the 
interaction energy.\ \cite{FraserModel,WilliamsonModel,KentModel}

For diamond, we estimated the finite size correction to be 
$-8.99$ mhartree/atom, using a converged LDA calculation with 220 k-points
in the irreducible Brillouin zone. 
For rhombohedral graphite, incorporating $d$ symmetry in the basis set
(as described above), and using 
189 k-points in the LDA calculation, we found the finite size 
correction to be $-6.56$ mhartree/atom.
We also include the calculated zero-point phonon energies 
of diamond and graphite, which 
are $6.65$ and $6.10$ mhartree/atom, respectively.\ \cite{zpe}

Adding all these corrections to the calculated VMC energies 
(Table\ \ref{tab:enercorr}), we estimate the energies of the 
infinite solids to be 
$-5.71529 \pm 0.00014$ hartree/atom for diamond 
and $-5.71341 \pm 0.00014$ hartree/atom for rhombohedral graphite. 
This appears to indicate that rhombohedral graphite is less stable than
diamond. 
However, given the approximation of using LDA finite size corrections, 
we might expect a systematic error of the order of 2 mhartree/atom 
in each of these results. 
This indicates that at the VMC level, the solids diamond and
rhombohedral graphite have very similar total energies.
We note that in the atomic pseudopotential used in the calculations presented
here $p$ and higher angular momentum scattering are all included in the
local potential. It is possible that the use of a separate $d$ 
pseudopotential might slightly affect the relative energies in both systems.

In order to determine the cohesive energy of a solid, we should subtract
the energy per atom of the solid from the energy of the isolated atom,
$E_c = E_a - E_s$.
However, when using approximate eigenfunctions, a reasonable estimate of
the cohesive energy is obtainable only by subtracting the energies estimated
using similar trial wave functions.
VMC energies are available for the carbon atom where the trial wave function
is of the Jastrow-Slater form.\ \cite{1bodyEFO} 
The orbitals of the Slater determinant are
optimized using energy minimization and a sophisticated Jastrow factor
is optimized using variance minimization, yielding a VMC energy for the atom
of $-5.4372 \pm 0.0001$ mhartree. Using this energy, we find the cohesive
energies to be $0.2781 \pm 0.0002$ hartree/atom for diamond and
$0.2762 \pm 0.0002$ hartree/atom for graphite.
We regard this atomic trial wave function to be close in form and accuracy to  
our solid trial wave function. However, to remain consistent with the inclusion
of $d$ symmetry in the basis set of our LDA calculations, we could also
refer to a multiconfiguration trial wave function for the carbon atom
which includes $d$ excitations. The VMC energy of the atom, using this
wave function, is $-5.45061 \pm 0.00002$ hartree,\ \cite{1bodyEFO}
leading to estimates of the cohesive
energies of diamond and rhombohedral graphite of 
$0.2647 \pm 0.0002$ hartree/atom and $0.2628 \pm 0.0002$ hartree/atom, 
respectively.
The experimental values are
0.271 hartree/atom for diamond, and 0.272 hartree/atom for 
graphite.\ \cite{BerkeleyExp}

Considering the significant gain in energy obtained
using an inhomogeneous Jastrow factor rather than a homogeneous form, one
might wonder how much the difference between VMC and DMC energies 
has been reduced.
Using Jastrow correlation factors which include one-body terms and homogeneous
two-body terms, similar in form to that used in Sec.\ \ref{Sec.HomoJ},
for bulk carbon\ \cite{Kentprivate} and silicon,\ \cite{LiSilicon}
typically leads to VMC correlation energies that are approximately 90\%
of the corresponding DMC correlation energies, i.e.,
\begin{eqnarray*}
  \frac{E_{\rm VMC} - E_{\rm HF}}{E_{\rm DMC} - E_{\rm HF}} \approx 0.90~,
\end{eqnarray*}
where $E_{\rm VMC}$ is the energy calculated using VMC, 
$E_{\rm DMC}$ is the energy calculated using DMC, and
$E_{\rm HF}$ is the Hartree-Fock energy.
($E_{\rm HF}$ has been estimated using just the LDA Slater 
determinant as trial wave function.\ \cite{FahyWang})
We have observed that including inhomogeneity in the correlation factor
leads to a gain in energy of approximately 10 mhartree/atom in the 
largest simulations of both diamond and graphite.
If we estimate that the VMC correlation energy, determined using 
our optimized wave function with homogeneous correlation terms,
obtains 90\% of $E_{\rm DMC} - E_{\rm HF}$,
then using the optimized wave function with inhomogeneous 
correlation terms obtains 96\% of the DMC correlation energy. 
This would indicate that the energy difference between  
VMC and DMC for these solids has been decreased by 60\%. 

It is important to emphasize that, whatever physical conclusions we would
like to draw from these calculations, our principal aim has been
to optimize the trial wave function for a given 
Hamiltonian, such that the expectation value of the
total energy is minimized. This aim has been
achieved for all the systems studied.

% ------------------------------------------------------------------------
\subsection{Correlation Factors}
\label{Sec.CorrelationFactors}
% ------------------------------------------------------------------------

Diamond, with a relatively homogeneous and isotropic electron charge
density, exhibits  an approximately homogeneous, 
isotropic Jastrow correlation function $u$ (Fig.\ \ref{fig:diamuopt}).
At large electron separations (beyond 6 a.u.) 
we see that the electron correlation factor $u$
in diamond is well approximated by homogeneous and isotropic functions, 
as all the curves plotted are quite similar. 
Only slight deviations from homogeneity exist at short and 
intermediate electron separations.
This inhomogeneity may be seen by comparing $u$ plotted with its fixed
coordinate at different points: $AA'$ and $AA''$ are quite similar at short
range, but clearly distinct from $BB'$ and $BB''$ in the same region.
This inhomogeneity may have significant effects on the short-range
pair-correlation functions calculated for diamond-like systems
using VMC methods.\ \cite{Fahycorrelation,Hoodcorrelation1,Hoodcorrelation2}

Graphite, with clear regions of high electron charge density and well-defined
regions of very low electron charge density between its layers, 
is a highly inhomogeneous and anisotropic structure. 
This is borne out in Fig.\ \ref{fig:graphuopt},
where the function $u$ differs considerably 
in various regions and in various directions.
At short-range, $u$ is surprisingly homogeneous in comparison with diamond. 
At intermediate separations the function displays
both inhomogeneous and anisotropic behaviour. Given that the
layer separation in these simulations is 6.33 a.u., this indicates that
inhomogeneous correlation between adjacent layers in the system is
not insignificant. 
This may prove important for van der Waals interactions
between the layers in graphite.

At long range, the anisotropy of the graphite correlation
factor is clearly shown in Fig. 13, where the correlation factors for 
electron separation vectors ${\bf r'-r}$ lying parallel to the graphite 
planes ($AA'$ and $BB'$) are distinctly different from those with
the separation vector perpendicular to the planes ($AA''$ and $BB''$).
Inhomogeneity (i.e. an explicit dependence on the position ${\bf r}$
of the first electron) is displayed at long range in the differences
between the function along $AA'$ and $BB'$. We might expect this,
given that the line AA' lies within the graphite planes, where the 
charge density is concentrated, whereas $BB'$ lies in the very low
charge density region between the planes [see Fig.\ \ref{fig:superlattice}(b)].
On the other hand, when the separation vector is oriented
perpendicular to the graphite planes ($AA''$ and $BB''$), 
inhomogeneous effects are substantially smaller at long range.

% ------------------------------------------------------------------------
\subsection{Computational Details}
\label{Sec.ComputationalDetails}
% ------------------------------------------------------------------------

In order to reduce the complexity of the physical 
results in Sec.\ \ref{Sec.Results}, 
some computational details of the method were not discussed.
We present and discuss some of these details in this section.

(1) This iterative method is trivially parallelizable. 
To determine the expectation values required to construct the predictor,
Monte Carlo sampling may be performed independently on many workstations
and the results combined.
To obtain total energies of the accuracy presented in this paper requires
$\sim 10^5$ Monte Carlo samples. However, for optimization of the wave function
using our method, this amount of sampling is also sufficient for 
accurate estimations of the expectation values required to construct 
the predictor (Sec.\ \ref{Sec.PredictorNumer}).

The extra time required to accumulate the various 
contributions to the predictor 
represents less than $5\%$ of the
computational time needed to calculate each energy sample. 
There are two reasons for this:
Firstly, because we have expressed the variational components of the
Jastrow factor as linear combinations of the operators ${\cal O}_m$ 
of Sec.\ \ref{Sec.EnerMin}. These ${\cal O}_m$ need only be evaluated
once, for each electron configuration, in order to construct both the
Jastrow factor and the predictor.
Secondly, and more importantly, since we must calculate the 
total energy as a sum of various
contributions defined by the Hamiltonian of the system,
all the energy contributions needed to construct the predictor 
are already available, either directly or by some simple manipulation.
It is also important to note that, 
despite the complex form of the Jastrow factor outlined in 
Sec.\ \ref{Sec.wavefn}, the amount of computational time spent evaluating it
is still less than or equal to that spent evaluating the Slater determinant,
for a given electron configuration.

The largest calculations presented in this paper were performed on
a Beowulf cluster of fifteen 500 MHz dual-processor workstations.
For a 5 iteration optimization using $10^5$ samples per iteration,
these calculations took about 50 hours on this cluster.
However, perhaps half this amount of sampling would have produced
comparable results [see (3) below]. 
The required memory for storing all
the expectation values necessary for this calculation 
was approximately 20 MB. 

(2) The Newton-Raphson method (see Appendix\ \ref{App.NewtonRaphson}), while
quadratically convergent near a root of a multidimensional system,
does possess some convergence problems far from the root. In the
calculations presented in this paper, we found that below a certain
minimum amount of sampling, the noise in the estimated expectation values
used to construct the predictor caused divergence of the solution to
Eq.\ \ref{eq:generateparam} using
the Newton-Raphson method. This problem might be solved through the use
of a more robust root-finding method for the predictor function.

\begin{figure}
\begin{center}
  \leavevmode
  \includegraphics[angle=-90,width=8cm]{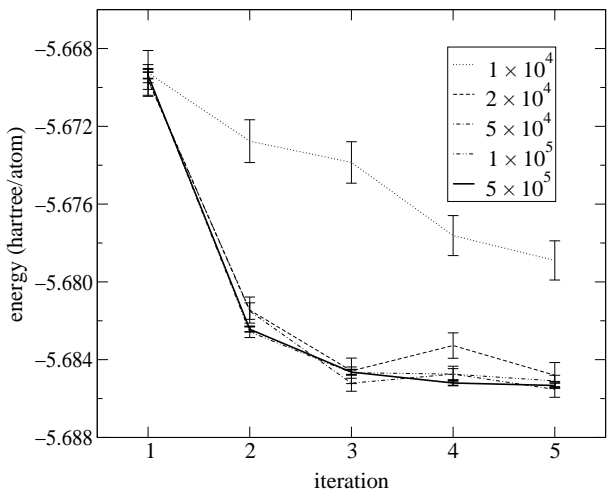}
  \caption{
    Total graphite energy in hartree/atom versus iteration number
    for a $2 \times 2 \times 2$ simulation region. 
    The number of Monte Carlo samples per iteration ranges from
    $10^4$ to $5 \times 10^5$.
  }
  \label{fig:Esamp}
\end{center}
\end{figure}

\begin{figure}
\begin{center}
  \leavevmode
  \includegraphics[angle=-90,width=8cm]{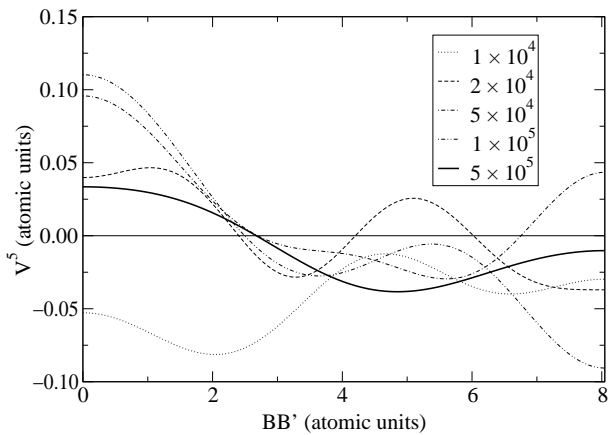}
  \caption{
    The reconstructed two-body function $V^5$, versus electron separation $r$
    along the line segment $BB'$ in graphite, from the fifth iteration
    of each of the calculations in Fig.\ \ref{fig:Esamp}.  
  }
  \label{fig:Vsamp}
\end{center}
\end{figure}

(3) From the analysis in Appendix\ \ref{App.ConvMeth}, we see that 
noise in the iterative method comes from the finite sampling used to
estimate those expectation values (listed in Sec.\ \ref{Sec.PredictorNumer}) 
required to construct the predictor.
The effect of this noise on the wave function accuracy is not clear.
To reduce the computational cost of these calculations we would prefer
to do the least amount of sampling necessary to produce the required results.

Figure\ \ref{fig:Esamp} illustrates the effect of various amounts of 
sampling on the convergence of the total energy in the optimization of the
$2 \times 2 \times 2$ simulation of rhombohedral graphite.
At $10^4$ samples per iteration, the estimation of the required expectation
values, during the first iteration of our method, is too crude to 
produce a convergent root of the predictor using
the Newton-Raphson method. This leads to
a wave function, used in the second iteration, with many noisy parameters
which are more difficult to optimize, as the convergence of the energy
shows. However, for sampling involving $2 \times 10^4$ 
samples per iteration, or more, 
we see that the convergence of the total energy is identical
(within statistical accuracy). Comparison of the optimized wave function
parameters also reveals only small differences, 
indicating that the only means of determining the true benefits of
more sampling is by examination of the fitted coefficients
$V_m(\mbox{\boldmath $\alpha$}^n)$ at each iteration $n$. 

In Fig.\ \ref{fig:Vsamp} we see, from the reconstruction 
$V^5({\bf r},{\bf r'})$ of the fitted coefficients for two-body optimization
at the fifth and final iteration of the method,
that the most sampling ($5 \times 10^5$ samples per iteration) 
reduces the magnitude of the Euler-Lagrange derivatives the most,
indicating that these wave function parameters are the most accurate. 
However, for practical purposes, there is little 
distinction between the accuracy of the wave function once we increase
the sampling beyond $2 \times 10^4$ samples per iteration.
Thus, the optimizations presented in Sec.\ \ref{Sec.Results} may have
used more computational time than was strictly necessary. 
However, more testing is required to determine 
the minimum amount of sampling as a function of system size.

% ------------------------------------------------------------------------
\section{Conclusion}
\label{Sec.Conclusion}
% ------------------------------------------------------------------------

We have developed a generalized form of electron correlation factor for 
trial many-body wave functions of electrons in periodic solids. 
This form allows us to represent fully inhomogeneous 
electron correlation in real physical periodic systems.
It is computationally efficient to evaluate, 
since the electron cusp, which we express as a homogeneous correlation factor,
is separated from the fully inhomogeneous form.

We have also developed a rapidly convergent iterative method 
for the optimization of all variational parameters in these
wave functions, minimizing the total energy of the given system.
It uses the accurate techniques of quantum Monte Carlo sampling to achieve
this optimization and has allowed new insights into the form of many-electron
correlation in systems with highly inhomogeneous charge densities.
We estimate the difference in energies calculated using the optimal
inhomogeneous two-body correlation factor and the optimal homogeneous
correlation factor to be approximately 60\% of the difference between
the DMC and VMC energies obtained with homogeneous 2-body correlation
terms.

In diamond, the optimal correlation factor is approximately homogeneous
and isotropic, with some inhomogeneity at short- and intermediate-range
electron separations. 
This is consistent with its comparatively homogeneous and isotropic
electron charge density. 
Graphite has an optimal correlation factor which is quite homogeneous
at short-range electron separation, but is significantly inhomogeneous
and anisotropic at intermediate- and long-range electron separations,
as one might expect from its highly inhomogeneous and anisotropic electron
charge density.
Nevertheless, it is remarkable that despite very large inhomogeneity in
the electron pair-correlation function, found by previous 
authors,\ \cite{Fahycorrelation,Hoodcorrelation1,Hoodcorrelation2} 
the ideal inhomogeneous
Jastrow two-body term, calculated for the first time here for diamond and
graphite, displays relatively small inhomogeneity. Whether this conclusion
can be extended to other systems (e.g. involving strongly correlated
$d$-electrons) remains an open question.

% ------------------------------------------------------------------------
\begin{acknowledgments}
% ------------------------------------------------------------------------

This work has been supported by Enterprise Ireland Grant Number SC/98/748
and by the Irish Higher Education Authority. The work of S.F. at the
National Microelectronics Research Centre is supported by Science
Foundation Ireland.

\end{acknowledgments}

% ------------------------------------------------------------------------
\appendix
% ------------------------------------------------------------------------

% ------------------------------------------------------------------------
\section{Derivative of observables with respect to wave function
         parameters}
\label{App.deriv}
% ------------------------------------------------------------------------

The derivative of the expectation value of an observable $A$,
with respect to a parameter $\alpha_m \in \mbox{\boldmath $\alpha$}$ 
of the parameterized
wave function $\Psi(\mbox{\boldmath $\alpha$})$, may be written as 
\begin{eqnarray*}
  \frac{\partial \langle A \rangle}{\partial \alpha_m} =
  \frac{\partial}{\partial \alpha_m} 
  \frac{\langle \Psi | A | \Psi \rangle}
       {\langle \Psi | \Psi \rangle}.
\end{eqnarray*}
We assume that $A$ is independent of the parameters $\mbox{\boldmath $\alpha$}$.
Also, for a time independent system, we may in general
express $\Psi$ as a real function.
Differentiating, we find that 
\begin{eqnarray*}
  \frac{\partial}{\partial \alpha_m} \langle \Psi | \Psi \rangle
  & = & 2 \langle \Psi | {\cal O}_m | \Psi \rangle
  \\ 
  \frac{\partial}{\partial \alpha_m} \langle \Psi | A | \Psi \rangle
  & = & 2 \langle \Psi | A {\cal O}_m | \Psi \rangle
\end{eqnarray*}
where we define the operator associated with variations of $\alpha_m$ as
${\cal O}_m \equiv \partial / \partial \alpha_m$, with local value
\begin{eqnarray*}
  {\cal O}_m({\bf R}) \equiv \frac{1}{\Psi({\bf R})} 
                \frac{\partial \Psi({\bf R})}{\partial \alpha_m}
  = \frac{\partial}{\partial \alpha_m} \ln \Psi({\bf R}) ~,
\end{eqnarray*}
for the many-body configuration ${\bf R}$.

Therefore, we may express the derivative of $\langle A \rangle$ as
\begin{eqnarray*}
  \frac{1}{2} \frac{\partial \langle A \rangle}{\partial \alpha_m} 
  & = &
  \frac{\langle \Psi | A {\cal O}_m | \Psi \rangle}{\langle \Psi | \Psi \rangle}
  - \frac{\langle \Psi | A | \Psi \rangle}{\langle \Psi | \Psi \rangle} 
    \frac{\langle \Psi | {\cal O}_m | \Psi \rangle}{\langle \Psi | \Psi \rangle}
  \\
  & = & \langle A {\cal O}_m \rangle 
      - \langle A \rangle \langle {\cal O}_m \rangle
  \\
  & = & \langle \Delta \! A \Delta \! {\cal O}_m \rangle,
\end{eqnarray*}
where 
$\Delta \! A({\bf R}) = A({\bf R}) - \langle A \rangle$ and
$\Delta \! {\cal O}_m({\bf R}) = 
  {\cal O}_m({\bf R}) - \langle {\cal O}_m \rangle$.

% ------------------------------------------------------------------------
\section{Least-squares fitting}
\label{App.leastsq}
% ------------------------------------------------------------------------

Our task is to minimize the integral
\begin{eqnarray*}
  \chi^2 = \langle \Psi | \left\{ {\cal H} - E_0 - 
                          \sum_{k} V_k {\cal O}_k \right\}^2 | \Psi \rangle
\end{eqnarray*}
by choosing the appropriate parameters $E_0$ and $\{ V_k \}$. We note first
that, at the minimum, 
\begin{eqnarray*}
  E_0 = \langle E \rangle - \sum_k V_k \langle {\cal O}_k \rangle \ ,
\end{eqnarray*}
where $\langle E \rangle$ is the expectation value of the total energy
and $\langle {\cal O}_k \rangle$
is the expectation value of the operator ${\cal O}_k$.
To fulfill the minimization, we must set all the remaining first derivatives
of $\chi^2$ to zero, i.e., 
\begin{eqnarray*}
  \frac{\partial \chi^2}{\partial V_l} = - 2 \langle \Psi | 
  \left\{ {\cal H} - E_0 - \sum_{k} V_k {\cal O}_k \right\}
  {\cal O}_l | \Psi \rangle = 0 \ .
\end{eqnarray*}
for each $l$.
Upon substitution of the minimum value of $E_0$, this leads to
\begin{eqnarray*}
  \langle E \, {\cal O}_l \rangle - \langle E \rangle \langle {\cal O}_l \rangle
  =  \sum_k V_k [ \langle {\cal O}_k {\cal O}_l \rangle -
     \langle {\cal O}_k \rangle \langle {\cal O}_l \rangle ] \ .
\end{eqnarray*}
Since, for any operators ${a}$, ${b}$ we may say that
\begin{eqnarray*}
  \langle a \, b \, \rangle  - \langle a \rangle \langle b \, \rangle =
  \langle \Delta a \, \Delta b \, \rangle
\end{eqnarray*}
where $\Delta a = a - \langle a \rangle$, then we are lead to the conclusion
that the least squares fitting is equivalent to solving the linear
system
\begin{eqnarray*}
  \sum_k V_k \langle \Delta {\cal O}_k \, \Delta {\cal O}_l \rangle
  =  \langle \Delta E \, \Delta {\cal O}_l \rangle
\end{eqnarray*}
for each $l$.

% ------------------------------------------------------------------------
\section{Newton-Raphson Method}
\label{App.NewtonRaphson}
% ------------------------------------------------------------------------

An integral part of the iterative procedure outlined in 
Sec.\ \ref{Sec.IterProc} is the determination of the parameters
$\mbox{\boldmath $\alpha$}$ 
that solve the system in Eq.\ \ref{eq:generateparam}.
The determination of the roots of any multidimensional function
can be troublesome. In all the optimizations presented in this
paper, the predictor function 
$V'_m(\mbox{\boldmath $\alpha$};\mbox{\boldmath $\alpha$}^0)$ is a
quadratic function of $\mbox{\boldmath $\alpha$}$, whose coefficients
are determined analytically, or by numerical fitting at the point
$\mbox{\boldmath $\alpha$}^0$. We ignore the implicit dependence of
$V'_m$ on $\mbox{\boldmath $\alpha$}^0$ for the purpose of finding a root,
and solve the system $V'_m(\mbox{\boldmath $\alpha$})=0$ for all $m$.

We use the Newton-Raphson
method\ \cite{NumRec} to determine the roots. 
This is an iterative method of improving successive guesses for a 
root of a function. It involves
the computation of the function $V'_m$ and its Jacobian matrix of derivatives
with respect to $\mbox{\boldmath $\alpha$}$, 
at each guess. 
The iterations continue until convergence of the solution is achieved
within a predefined tolerance.

The success of the Newton-Raphson method for multi-dimensional systems
relies heavily on the proximity of the initial guess to the root we seek. 
For this reason, the initial guess chosen is normally the parameter set
of the previous iteration of the procedure outlined in Sec.\ \ref{Sec.IterProc}.
To find the first set of parameters, by finding the roots of the
analytic predictor $S_m$, we require a good initial guess of the root.
Since the one-body parameters $\chi_{\bf G}$ are expected to be small
by construction, 
an initial guess of zero for all ${\bf G}$ was found to be sufficient 
to produce a convergent solution to the first application of the
Newton-Raphson method.

For the two-body problem, we rescale the variables $u_{\bf q q'}$ to
improve the convergence of the root-finding method.
According to the RPA,\ \cite{BohmPines} the long-range behaviour of
the $u$ function should take the form $u(r)= 1/\omega_p r$, where
the plasma frequency for a homogeneous system with electron charge
density $n$ is $\omega_p = \sqrt{4 \pi n}$. 
The charge density $n$ is determined in the simulation region to be
$N/\Omega$, for $N$ the number of electrons in the simulation volume
$\Omega$. Therefore, for small ${\bf q}$
in Fourier space, $u$ behaves like 
\begin{eqnarray*}
  u({\bf q}) = \frac{4 \pi}{\Omega \omega_p q^2} 
             = \omega_p \frac{1}{N q^2}~.
\end{eqnarray*}
This indicates large relative differences between values of $u({\bf q})$ for
small ${\bf q}$. Therefore, in inhomogeneous systems, it would be 
appropriate to rescale the parameters
$u_{\bf q q'}$ by multiplying by $N |{\bf q}| |{\bf q'}|$, thus rendering
all the variables of the same order of magnitude as the plasma frequency. 
This leads to a less pathological numerical problem for the Newton-Raphson
method. Appropriate scaling must also be applied to the predictor function
$V'_{\bf q q'}$ in Sec.\ \ref{Sec.EnerMin}.
For the initial analytic guess of the roots of $S_{\bf q q'}$
we use the homogeneous solution for $u_{\bf q q'}$ outlined in 
Eq.\ \ref{eq:homosoln}.

% ------------------------------------------------------------------------
\section{One-body Energy Contributions}
\label{App.1bEnerCon}
% ------------------------------------------------------------------------

The variational part of $\Psi$ associated with one-body terms is the
Jastrow factor $J_{\rm 1b}$ (Eq.\ \ref{eq:1bJast}) with parameters 
$\mbox{\boldmath $\chi$} = \{\chi_{\bf G}\}$.
We expand the energy contributions 
$\epsilon^{(1)}$ and $\epsilon^{(2)}$ defined in 
Sec.\ \ref{Sec.EnergyPartition}.
For $J_{\rm 1b}$, we have that
\begin{eqnarray*}
  \epsilon^{(1)} &\equiv& -\frac{1}{2} \sum_i \frac{\nabla_i^2 J_{\rm 1b}}
                                                   {J_{\rm 1b}}
  \\
  &=& -\frac{1}{2} \sum_i ( \nabla_i^2 \ln J_{\rm 1b} +
      | \mbox{\boldmath $\nabla$}_i \ln J_{\rm 1b} |^2 )
  \\
  &=&\frac{1}{2} \sum_{\bf G} \chi_{\bf G} G^2 \Delta \rho_{\bf G}^*
  \\
  && +\frac{1}{2} \sum_{\bf G G'} \chi_{\bf G-G'} ({\bf G-G'}){\bf \cdot G'}
                                \chi_{\bf G'} \Delta \rho_{\bf G}^*
    + \mbox{constant}~.
\end{eqnarray*}
The constant terms are not required, so we ignore them.

The energy contribution $\epsilon^{(2)}$
cannot be expanded analytically as a linear combination of the functions
$\Delta \rho_{\bf G}^*$ since $\Psi' \equiv \Psi/J_{\rm 1b}$ is not
explicitly a function of these coordinates. 
\begin{eqnarray*}
  \epsilon^{(2)} &\equiv& - \sum_i \frac{\mbox{\boldmath $\nabla$}_i J_{\rm 1b}}
                                 {J_{\rm 1b}} \mbox{\boldmath $\cdot$}
                                 \frac{\mbox{\boldmath $\nabla$}_i \Psi'}
                                 {\Psi'}
  \\
  &=& - \sum_{\bf G} \chi_{\bf G}
  \sum_i \mbox{\boldmath $\nabla$}_i \Delta \! \rho_{\bf G}^*
         \mbox{\boldmath $\cdot$}
         \frac{\mbox{\boldmath $\nabla$}_i \Psi'}{\Psi'} 
\end{eqnarray*}

However, $\epsilon^{(2)}$ is explicitly linear in the parameters
$\mbox{\boldmath $\chi$}$, with derivatives
\begin{eqnarray*}
  \frac{\partial \epsilon^{(2)}}{\partial \chi_{\bf G}} =
  - \sum_i \mbox{\boldmath $\nabla$}_i \Delta \! \rho_{\bf G}^*
         \mbox{\boldmath $\cdot$}
         \frac{\mbox{\boldmath $\nabla$}_i \Psi'}{\Psi'} ~,
\end{eqnarray*}
which are independent of $\mbox{\boldmath $\chi$}$.

% ------------------------------------------------------------------------
\section{Two-body Energy Contributions}
\label{App.2bEnerCon}
% ------------------------------------------------------------------------

The variational part of $\Psi$ associated with two-body energy contributions
is the inhomogeneous Jastrow factor $J_{\rm ih}$ of Eq.\ \ref{eq:inhomJast},
with parameters ${\bf u} = \{ u_{\bf q q'} \}$.
(For periodic systems we use only $u_{{\bf q+G},{\bf q+G'}}$.)
The energy contributions $\epsilon^{(i)}$ of Sec.\ \ref{Sec.EnergyPartition}
are expanded here.
The contribution $\epsilon^{(1)}$ is dependent only on the form of $J_{\rm ih}$
and is expanded as
\begin{eqnarray*}
  \epsilon^{(1)} \equiv - \frac{1}{2} \sum_i 
    \frac{\nabla_i^2 J_{\rm ih}}{J_{\rm ih}} 
    = - \frac{1}{2} \sum_i ( \nabla_i^2 \ln J_{\rm ih} +
      | \mbox{\boldmath $\nabla$}_i \ln J_{\rm ih} |^2 )~.
\end{eqnarray*}
We find that
\begin{eqnarray*}
  \lefteqn{- \frac{1}{2} \sum_i \nabla_i^2 \ln J_{\rm ih} =
           - \frac{1}{2} \sum_{\bf q q'} u_{\bf q q'} (q^2 + q'^2) 
             P_{\bf q q'}}
  \hspace{0.5cm}
  \\
  & & -\sum_{\bf q} \left[ \frac{N-1}{N} \sum_{\bf q'} u_{\bf q q'} 
                            q^2 \langle \rho_{\bf q'} \rangle \right]
                            \Delta \rho_{\bf q}^*
      + \mbox{constant}~. 
\end{eqnarray*}
We retain only the linear combination of the
functions $P_{\bf q q'}$. The constant terms we may ignore, 
and the one-body terms (linear combinations of $\Delta \rho_{\bf q}^*$) 
we assume are compensated by terms in the one-body Jastrow $J_{\rm 1b}$.

If we ignore the removal of one-body terms from the function
$P_{\bf q q'} = 
 (\Delta \rho_{\bf q}^{\mbox{}} \Delta \rho_{\bf q'}^*)_{[i \neq j]}$
and consider using the function
$\Delta \rho_{\bf q}^{\mbox{}} \Delta \rho_{\bf q'}^*$ instead, 
then we find that
\begin{eqnarray}
  \label{eq:fullquad}
  \lefteqn{- \frac{1}{2} \sum_i | \mbox{\boldmath $\nabla$}_i 
             \ln J_{\rm ih} |^2 =} \hspace{0.5cm} \\ 
  & & -2 \sum_{\bf q q'} \sum_{\bf k k'}
      u_{\bf q k} ({\bf k \cdot k'}) u_{\bf k' q'}
      \rho_{\bf k' - k} 
      \Delta \rho_{\bf q}^{\mbox{}} \Delta \rho_{\bf q'}^* 
      + \cdots ~,
  \nonumber
\end{eqnarray}
where we have ignored constant and one-body terms in the expansion. 
The product 
$\rho_{\bf k' - k} \Delta \rho_{\bf q}^{\mbox{}} \Delta \rho_{\bf q'}^*$
contains both two- and three-body terms, since we may 
rewrite $\rho_{\bf k' - k}$ as 
$\Delta \rho_{\bf k'-k} + \langle \rho_{\bf k'-k} \rangle$.
We intend here to remove two-body fluctuations and regard three-body
fluctuations as much less significant, so we retain only the
two-body term\ \footnote{ 
        This approximation is related to the RPA of 
        Bohm and Pines\protect\ \protect\cite{BohmPines}
        and is used in the work of Bevan\protect\ \protect\cite{Bevan}, 
        Gaudoin \emph{et al.}\protect\ \protect\cite{Gaudoin} 
        and Ceperley.\protect\ \protect\cite{Ceperley}
}
$\langle \rho_{\bf k' - k} \rangle \Delta \rho_{\bf q}^{\mbox{}} 
 \Delta \rho_{\bf q'}^*$
from the charge fluctuation products in Eq.\ \ref{eq:fullquad}, i.e., 
\begin{eqnarray*}
  -2 \sum_{\bf q q'} \sum_{\bf k k'}
    u_{\bf q k} ({\bf k \cdot k'}) u_{\bf k' q'}
    \langle \rho_{\bf k' - k} \rangle 
    \Delta \rho_{\bf q}^{\mbox{}} \Delta \rho_{\bf q'}^* 
    + \cdots ~.
\end{eqnarray*}
Now we make the assumption that removing one-body terms from this
expression , i.e., replacing 
$\Delta \rho_{\bf q}^{\mbox{}} \Delta \rho_{\bf q'}^*$
with $P_{\bf q q'}$ is a good approximation,
and obtain the expression for $v_{\bf q q'}^{(1)}({\bf u})$
given in Sec.\ \ref{Sec.PredictorAnalApprox.J2b}. 

The contribution $\epsilon^{(2)}$ cannot be expanded
analytically in the basis of fluctuation functions $P_{\bf q q'}$,
However, it is clear that $\epsilon^{(2)}$ is 
linear in ${\bf u}$, since, for $\Psi' \equiv \Psi/J_{\rm ih}$,
\begin{eqnarray*}
  \epsilon^{(2)} = \sum_{\bf q q'} u_{\bf q q'} 
                \sum_i {\bf \nabla}_i P_{\bf q q'}
                {\bf \cdot} \frac{{\bf \nabla}_i \Psi'}{\Psi'} ~,
\end{eqnarray*}
and has first derivatives
\begin{eqnarray*}
  \frac{\partial \epsilon^{(2)}}{\partial u_{\bf q q'}} =
    \sum_i {\bf \nabla}_i P_{\bf q q'}
           {\bf \cdot} \frac{{\bf \nabla}_i \Psi'}{\Psi'}~.
\end{eqnarray*}
which are independent of the parameters ${\bf u}$.

The final energy contribution $\epsilon^{(3)}$ we attempt to express
analytically in terms of the two-body fluctuation functions $P_{\bf q q'}$.
If $\Psi' \approx D$, 
the LDA Slater determinant, then the sum of contributions 
from the external potential and the kinetic energy term
$-(1/2) \sum_i \nabla_i^2 \Psi' / \Psi'$ is a one-body contribution
defined by the Kohn-Sham Hamiltonian,\ \cite{KohnSham} since 
\begin{eqnarray*}
  \sum_i [ - \frac{1}{2} \frac{\nabla_i^2 D}{D} + V_{\rm ext}({\bf r}_i) ]
  = \sum_i [ \epsilon^{\rm KS}_i 
             - V_{\rm H}({\bf r}_i) - V_{\rm xc}({\bf r}_i) ]~,
\end{eqnarray*}
where $\epsilon^{\rm KS}_i$ are the Kohn-Sham eigenvalues, 
$V_{\rm H}$ is the Hartree potential and 
$V_{\rm xc}$ is the exchange and correlation potential.
Therefore, the two-body contribution of these terms is approximately zero.
We are left with the contribution of the electron-electron
interaction $V$.

For a two-body potential $V({\bf r}, {\bf r'})$, we may
expand the sum over electron pairs as
\begin{eqnarray*}
  \sum_{i < j} V({\bf r}_i, {\bf r}_j) &=& 
    \frac{1}{2} \sum_{\bf q q'} V_{\bf q q'} 
    \sum_{i \neq j} e^{-i{\bf q \cdot r}_i} e^{i{\bf q' \cdot r}_j}
    \\
    &=& \frac{1}{2} \sum_{\bf q} V_{\bf q q'} 
      (\rho_{\bf q} \rho_{\bf q'}^*)_{[i \neq j]}~,
\end{eqnarray*}
for Fourier coefficients $V_{\bf q q'}$.
In terms of the fluctuation functions $P_{\bf q q'}$, this may be rewritten as
\begin{eqnarray*}
  \lefteqn{\sum_{i < j} V({\bf r}_i, {\bf r}_j) = 
    \frac{1}{2} \sum_{\bf q q'} V_{\bf q q'} P_{\bf q q'}}
  \hspace{0.5cm} \\
  & & - \sum_{\bf q} \left[ \frac{N-1}{N} \sum_{\bf q'} V_{\bf q q'} 
    \langle \rho_{\bf q'} \rangle \right]
    \Delta \rho_{\bf q}^*
  + ~\mbox{constant}~.
\end{eqnarray*}
(We assume that $V({\bf r}, {\bf r'})$ possesses exchange symmetry, 
so that
$V_{\bf q q'} = V_{\bf -q' -q}$.)
Therefore, both one- and two-body fluctuations arise from a two-body
potential in the ``charge fluctuation'' coordinate system.    
Note that the one-body fluctuations are expressible in terms of the
Hartree potential, since
\begin{eqnarray*}
  V_{\rm H}({\bf q}) = \sum_{\bf q'} V_{\bf q q'} 
                       \langle \rho_{\bf q'} \rangle~.
\end{eqnarray*}

If the two-body potential is homogeneous, i.e.
$V({\bf r}, {\bf r'}) = V({\bf r-r'})$, then we may simplify
the fluctuations since 
$V_{\bf q q'} = V_{\bf q}^* \delta({\bf q-q'})$, 
where the Fourier transform of the homogeneous function $V({\bf r})$ is
\begin{eqnarray*}
  V_{\bf q} \equiv \frac{1}{\Omega} \int_{\Omega} 
                   V({\bf r}) e^{-i {\bf q \cdot r}} d{\bf r}~,
\end{eqnarray*}
for a system volume $\Omega$.
If $V({\bf r}) = V(-{\bf r})$ then $V_{\bf q}^* = V_{\bf q}$.
Therefore,
\begin{eqnarray}
  \label{eq:V2bodyhomo}
  \sum_{i < j} V(|{\bf r}_i-{\bf r}_j|) &=&
    \frac{1}{2} \sum_{\bf q} V_{\bf q} P_{\bf q q} \\
  &-& \frac{N-1}{N} \sum_{\bf q} V_{\bf q}
    \langle \rho_{\bf q} \rangle 
    \Delta \rho_{\bf q}^*
  + ~\mbox{constant}~.
  \nonumber
\end{eqnarray}
Again we ignore the one-body and constant terms in this context.

% ------------------------------------------------------------------------
\section{Consequences of using the short-range Jastrow}
\label{App.ShortJConseq}
% ------------------------------------------------------------------------

The short range Jastrow factor $J_{\rm sr}$ is constructed from the
pseudointeraction $V_{\rm ps}$ in the following way. For the 
isolated two-electron scattering problem, we may find an
eigenstate $\psi_0$ of the true two-electron Hamilatonian $h_0$
for a given energy eigenvalue $\epsilon$. Upon replacing the
true Coulomb interaction $V$ with a generated pseudointeraction 
$V_{\rm ps}$, we construct the modified Hamiltonian $h_{\rm ps}$
with eigenstate $\psi_{\rm ps}$ corresponding to $\epsilon$.
We construct $J_{\rm sr}$ such that
\begin{eqnarray*}
  \psi_0 = J_{\rm sr} \psi_{\rm ps}~.
\end{eqnarray*}

Now, in a many-electron environment, we know that using $J_{\rm sr}$
allows for good approximation of short-range correlations.\ \cite{Cusppaper}
We might imagine that for a many-electron system with Hamiltonian ${\cal H}$
and many-electron trial wave function 
$J_{\rm sr} \Psi_{\rm ps}$, 
the true energy eigenvalue may be well approximated by
\begin{eqnarray*}
  E \approx \frac{{\cal H} J_{\rm sr} \Psi_{\rm ps}}{J_{\rm sr} \Psi_{\rm ps}}
    \approx \frac{{\cal H}_{\rm ps} \Psi_{\rm ps}}{\Psi_{\rm ps}}~,
\end{eqnarray*}
where 
${\cal H}_{\rm ps} = {\cal H} - \sum_{i<j} [V(r_{ij})-V_{\rm ps}(r_{ij})]$. 
Now, disguising the true interaction $V$ with $V_{\rm ps} + (V-V_{\rm ps})$,
and given the transferability of $V_{\rm ps}$ over a wide range of energies,
we see that
\begin{eqnarray*}
  \lefteqn{\sum_{i<j} V_{\rm ps}(r_{ij}) \approx} \hspace{0.5cm} 
    \\
    & & - \frac{1}{2} \sum_i \frac{\nabla_i^2 J_{\rm sr}}{J_{\rm sr}}
    - \sum_i \frac{\mbox{\boldmath $\nabla$}_i J_{\rm sr}}{J_{\rm sr}} \cdot
             \frac{\mbox{\boldmath $\nabla$}_i \Psi_{\rm ps}}{\Psi_{\rm ps}} 
    + \sum_{i<j} V(r_{ij})~.
\end{eqnarray*}
 
Dividing two-body correlation into short-range and inhomogeneous terms,
we use a Jastrow factor of the form $J_{\rm sr} J_{\rm ih}$.
The local energy determined using the trial wave function
$\Psi = J_{\rm sr} J_{\rm ih} \Psi'$, where 
$\Psi' \equiv \Psi / (J_{\rm sr} J_{\rm ih})$, may be expanded as
\begin{eqnarray*}
  E &= -&\frac{1}{2} \sum_i \frac{\nabla_i^2 J_{\rm sr}}{J_{\rm sr}}
      - \sum_i \frac{\mbox{\boldmath $\nabla$}_i J_{\rm sr}}{J_{\rm sr}} \cdot
               \frac{\mbox{\boldmath $\nabla$}_i \Psi/J_{\rm sr}}
                    {\Psi/J_{\rm sr}}
      + \sum_{i<j} V(r_{ij})
      \\
      &\hspace{1em}-&\frac{1}{2} \sum_i \frac{\nabla_i^2 J_{\rm ih}}{J_{\rm ih}}
      - \sum_i \frac{\mbox{\boldmath $\nabla$}_i J_{\rm ih}}{J_{\rm ih}} \cdot
               \frac{\mbox{\boldmath $\nabla$}_i \Psi'}{\Psi'}
      -\frac{1}{2} \sum_i \frac{\nabla_i^2 \Psi'}{\Psi'}
      \\
      &\hspace{1em}+& \sum_i V_{\rm ext}({\bf r}_i)
    \\
    &\approx \hspace{1em}& \sum_{i<j} V_{\rm ps}(r_{ij})
      -\frac{1}{2} \sum_i \frac{\nabla_i^2 J_{\rm ih}}{J_{\rm ih}}
      - \sum_i \frac{\mbox{\boldmath $\nabla$}_i J_{\rm ih}}{J_{\rm ih}} \cdot
               \frac{\mbox{\boldmath $\nabla$}_i \Psi'}{\Psi'}
      \\
      &\hspace{1em}-&\frac{1}{2} \sum_i \frac{\nabla_i^2 \Psi'}{\Psi'}   
      + \sum_i V_{\rm ext}({\bf r}_i)~.
\end{eqnarray*} 
For this reason, we use $V_{\rm ps}$ in the expansion of the local energy
for $J_{\rm ih}$ to implicitly include the short-range Jastrow factor
$J_{\rm sr}$.

% ------------------------------------------------------------------------
\section{Convergence of the method}
\label{App.ConvMeth}
% ------------------------------------------------------------------------

The convergence criterion $V_m(\mbox{\boldmath $\alpha$})=0$, is
numerically never exactly achieved.
Given that the predictor $V_m'$ contains
some terms determined by statistical fitting, finite sampling errors
exist, and this noise is passed on to the fitted parameters in $V_m'$
from Eqs.\ \ref{eq:lstsqsys} and\ \ref{eq:contributionlstsqsys}.

We consider ${\bf V}_n = \{ V_m(\mbox{\boldmath $\alpha$}^n) \}$, 
the fitted coefficients of the local energy at the 
$n^{\rm th}$ iteration of the method. 
The method may be regarded as an iterative map ${\bf M}$, such that
the coefficients are determined via ${\bf V}_{n+1} = {\bf M}( {\bf V}_n )$.
There are two sources of noise in ${\bf V}_{n+1}$:
(i) noise inherited from ${\bf V}_n$, which produced the parameters
    $\mbox{\boldmath $\alpha$}^{n+1}$, which were used to construct the
    wave function $\Psi(\mbox{\boldmath $\alpha$}^{n+1})$, with which we
    evaluated the expectation values used to calculate ${\bf V}_{n+1}$;
and (ii) finite sampling noise in the evaluation of the expectation values
         in Eqs.\ \ref{eq:lstsqsys} and\ \ref{eq:contributionlstsqsys}
         via Monte Carlo sampling.
Therefore, we associate a set of variances 
$\mbox{\boldmath $\sigma$}^2_n = 
 \{ \sigma^2_m ~\mbox{for each $m$ at step $n$} \}$,
arising from these two sources of noise,
to the coefficients ${\bf V}_{n}$. 
The variance due to finite sampling alone, at each step $n$, is
${\bf s}^2_n$ and we use the initial condition 
$\mbox{\boldmath $\sigma$}^2_1 = {\bf s}^2_1$.
This implies that the variance obeys the following
iterative map:
\begin{eqnarray*}
  \mbox{\boldmath $\sigma$}^2_{n+1} \approx {\bf s}^2_{n+1}
  + | \mbox{\boldmath $\lambda$}_n |^2 \mbox{\boldmath $\sigma$}^2_{n}~,
\end{eqnarray*}
where 
$\mbox{\boldmath $\lambda$}_n = 
  \mbox{\boldmath $\nabla$}_{\bf V} {\bf M}({\bf V}_{n})$.
For a convergent map ${\bf M}$, we are guaranteed that
$| \mbox{\boldmath $\lambda$}_n |^2 < 1$
at convergence.
$| \mbox{\boldmath $\lambda$}_n |$ is a measure of the convergence rate of
the map ${\bf M}$, with $| \mbox{\boldmath $\lambda$}_n | \approx 0$
implying fast convergence and $| \mbox{\boldmath $\lambda$}_n | \approx 1$
implying slow convergence. 
Note that if $| \mbox{\boldmath $\lambda$}_n | \ge 1$ the map is divergent. 

Therefore, if we regard ${\bf s}_n \approx {\bf s}$ and
$\mbox{\boldmath $\lambda$}_n \approx \mbox{\boldmath $\lambda$}$ for all $n$,
for constants ${\bf s}$ and $\mbox{\boldmath $\lambda$}$,
then the converged value of the variance in the fitted coefficients is
\begin{eqnarray}
  \mbox{\boldmath $\sigma$}^2_* = 
    \frac{1}{1-| \mbox{\boldmath $\lambda$} |^2} {\bf s}^2~.
  \label{eq:convvar}
\end{eqnarray}

We conclude that, given a convergent method ${\bf M}$, 
the presence of statistical noise 
does not lead to successively more ill-determined parameters 
$\mbox{\boldmath $\alpha$}$,
since their variance is also convergent. 

Note that Eq.\ \ref{eq:convvar} indicates that the variance in the
fitted coefficients $V_m(\mbox{\boldmath $\alpha$})$ is always
greater than or equal to the variance estimated using
finite sampling. However, our implementation of the method
represented by the map ${\bf M}$ indicates that 
$| \mbox{\boldmath $\lambda$} | \ll 1$, since we find that the majority
of the coefficients $V_m(\mbox{\boldmath $\alpha$})$ ultimately
end up with magnitudes approximately equal to their finite sampling errors,
signifying that statistically they are zero.

Therefore, the final conclusion to be drawn from Eq.\ \ref{eq:convvar} is
that the accuracy of our optimization method depends ultimately on the
finite sampling error. Therefore, increasing the computational workload,
by increasing the amount of sampling, will result in more accurate
optimizations of the wave function. 
This is demonstrated by the results in Sec.\ \ref{Sec.Discussion}.

\end{document}